\documentclass[fleqn,usenatbib]{mnras}

\usepackage{newtxtext,newtxmath}

\usepackage[T1]{fontenc}

\DeclareRobustCommand{\VAN}[3]{#2}
\let\VANthebibliography\thebibliography
\def\thebibliography{\DeclareRobustCommand{\VAN}[3]{##3}\VANthebibliography}

\usepackage{graphicx}	

\newcommand \msun {M$_\odot$}
\newcommand \kms {km\,s$^{-1}$}
\newcommand \dv {\mathrm{d}}
\newcommand \mh {[M/H]}
\newcommand \mhf {\left[\frac{M}{H}\right]}
\DeclareMathOperator{\erf}{erf}
\newcommand \af {[$\alpha$/Fe]}

\graphicspath{{./figures/}}

\title[A bottom-heavy IMF for blue-halo stars]{A bottom-heavy initial mass function for the likely-accreted blue-halo stars of the Milky Way}

\author[N. Hallakoun \& D. Maoz]{
Na'ama Hallakoun$^{1}$\thanks{E-mail: \href{mailto:naama.hallakoun@weizmann.ac.il}{naama.hallakoun@weizmann.ac.il} (NH)}
and Dan Maoz$^{2}$
\\
$^{1}$Department of particle physics and astrophysics, Weizmann Institute of Science, 7610001 Rehovot, Israel\\
$^{2}$School of Physics and Astronomy, Tel-Aviv University, Tel-Aviv 6997801, Israel
}

\date{Accepted XXX. Received YYY; in original form ZZZ}

\pubyear{2021}

\begin{document}
\label{firstpage}
\pagerange{\pageref{firstpage}--\pageref{lastpage}}
\maketitle

\begin{abstract}
We use \textit{Gaia} DR2 to measure the initial mass function (IMF) of stars within 250\,pc and masses in the range $0.2< m/M_\odot < 1.0$, separated according to kinematics and metallicity, as determined from \textit{Gaia} transverse velocity, $v_T$, and location on the Hertzsprung-Russell diagram (HRD). The predominant thin-disc population ($v_T<40$\,\kms) has an IMF similar to traditional (e.g. \citealt{Kroupa_2001}) stellar IMFs, with star numbers per mass interval $\dv N / \dv m$ described by a broken power law, $m^{-\alpha}$, and index $\alpha_\textrm{high}=2.03^{+0.14}_{-0.05}$ above $m\sim 0.5$, shallowing to $\alpha_\textrm{low}=1.34^{+0.11}_{-0.22}$ at $m\lesssim0.5$. Thick-disc stars (60\,\kms\ $<v_T<$ 150\,\kms) and stars belonging to the ``high-metallicity'' or ``red-sequence'' halo ($v_T > 100$\,\kms\ or $v_T>200$\,\kms, and located above the isochrone on the HRD with metallicity $\textrm{\mh}>-0.6$) have a somewhat steeper high-mass slope, $\alpha_\textrm{high}=2.35^{+0.97}_{-0.19}$ (and a similar low-mass slope $\alpha_\textrm{low}=1.14^{+0.42}_{-0.50}$). Halo stars from the ``blue sequence'', which are characterised by low-metallicity ($\textrm{\mh}<-0.6$), however, have a distinct, bottom-heavy IMF, well-described by a single power law with $\alpha=1.82^{+0.17}_{-0.14}$ over most of the mass range probed. The IMF of the low-metallicity halo is reminiscent of the Salpeter-like IMF that has been measured in massive early-type galaxies, a stellar population that, like Milky-Way halo stars, has a high ratio of $\alpha$ elements to iron, \af. Blue-sequence stars are likely the debris from accretion by the Milky Way, $\sim 10$\,Gyrs ago, of the Gaia-Enceladus dwarf galaxy, or similar events. These results hint at a distinct mode of star formation common to two ancient stellar populations---elliptical galaxies and galaxies possibly accreted early-on by ours.
\end{abstract}

\begin{keywords}
stars: luminosity function, mass function -- Hertzsprung--Russell and colour--magnitude diagrams -- Galaxy: stellar content -- solar neighbourhood -- methods: statistical -- stars: statistics
\end{keywords}



\section{Introduction}
\label{sec:intro}
It has long been a major goal in astrophysics to measure the initial mass-distribution function (IMF) with which stars form (e.g. \citealt{Miller_1979, Kroupa_2001, Chabrier_2003, Just_2010, Rybizki_2015, Mor_2019, Zonoozi_2019}; see \citealt{Bastian_2010} for a review). Countless theoretical predictions and interpretations of observations, in many astronomical sub-disciplines, rely on the assumption of an IMF. Even more important, perhaps, it has long been hoped that the observed IMF and its variations, if any, with cosmic time and star-forming environment, could serve as a fossil clue to the poorly understood process of star formation. The IMF might depend on the chemical composition and the structure of the molecular cloud which forms the stars, if fragmentation and competitive accretion are governing the stellar mass distribution \citep[e.g.][]{Larson_1978, Larson_1998, Jappsen_2005, Chabrier_2014}. Alternatively, it could be the stars themselves that determine their masses, through strong stellar outflows \citep{Adams_1996}.

While the IMF is defined as the stellar mass distribution of a single star formation event in an embedded cluster, the galaxy-wide IMF sums the contributions from all star-forming regions in the galaxy, over the galaxy lifetime \citep{Kroupa_2003}. If the IMF is indeed universal, the form of the galaxy-wide IMF should be similar to that of the canonical IMF. However, a galaxy-wide IMF that depends on the metallicity and star formation rate of the galaxy, could indicate a non-universal IMF \citep{Jerabkova_2018}.
Until the last decade, opinions seemed to favour the existence of a universal IMF, even if discord remained regarding the exact details of the IMF's functional form. More recently, however, evidence has been accumulating for IMF variations in at least some extragalactic environments, particularly in massive early-type galaxies.  
\citet{vanDokkum_2010}, \citet{Treu_2010}, \citet{Conroy_2012}, \citet{Cappellari_2012, Cappellari_2013}, \citet{Lyubenova_2016}, \citet{Conroy_2017}, \citet{Davis_2017}, \citet{Zhang_2018}, and others have deduced the existence of a ``bottom-heavy'' IMF in such galaxies: the IMF slope, rather than becoming shallower at stellar masses below $\sim 0.5$\,\msun, as in the most popular ``universal'' IMFs, continues with the steep ``Salpeter'' slope that characterises the higher stellar masses (or even steeper), down to the hydrogen-burning limit at $\sim 0.1$\,\msun. Most recently, there are indications that there exists an IMF-slope gradient with galactic radius, with the most bottom-heavy IMF in the central regions of elliptical galaxies \citep{Sarzi_2018, LaBarbera_2019}. \citet{Geha_2013} have measured a shallower, or ``bottom-light'' IMF in the $0.52-0.77$\,\msun\ range for two ultra-faint dwarf satellites of the Milky Way \citep[but see][]{ElBadry_2017}. There is a disagreement regarding the agents that drive these IMF variations (galaxy mass, metallicity, age, etc.).

Apart from their peculiar IMFs, the stellar atmospheres in massive elliptical galaxies display a high ratio of $\alpha$ elements to iron \af, compared to the stars in lower-mass and disc galaxies, and compared to the thin-disc stars in the Milky Way \citep[e.g.,][]{Edvardsson_1993, Venn_2004, Conroy_2014}. Since the bulk of the $\alpha$ elements are produced by core-collapse SNe (CC-SNe) from massive stars \citep{Matteucci_1986}, while iron is synthesized both in Type Ia supernovae (SNe Ia) and CC-SNe \citep[with roughly equal contributions to the universal iron budget, see][]{Maoz_2017}, a peculiar \af\ ratio is therefore suggestive of IMF variations at the high end of the IMF: a change in the high-mass IMF slope might change the resulting mix of different CC-SN types, thus changing the integrated \af\ from CC-SNe; and/or a change in the ratio of high-mass stars (that explode as CC-SNe) and intermediate-mass stars (some of whose white-dwarf descendants produce SNe Ia), could also affect \af. Indeed, recent measurements in galaxy clusters (which are dominated by massive early-type galaxies) have shown a high time-integrated production efficiency of SNe Ia \citep{Friedmann_2018, Freundlich_2021}. One potential way to enhance SN Ia production is via IMF modifications, namely by means of an excess of intermediate-mass stars, relative to solar-mass stars, as compared to the ratio in standard IMFs \citep[e.g.][]{Yan_2020}.

It has been known for some time that many of the stars belonging to the stellar-halo and thick-disc components of the Milky Way also display high \af\ \citep[e.g.][]{Walcher_2016, Ness_2016, Ho_2017}. \citet{Maoz_2017} have shown that a high time-integrated SN Ia production efficiency, as measured in galaxy clusters, is a possible way of reproducing the large drop seen in the \af\ of halo stars below the ``knee'' in diagrams of \af\ vs. iron abundance, [Fe/H]. Over the past three years, data from \textit{Gaia} Data Release 2 \citep[DR2;][]{Gaia_2016, GaiaDR2_2018}, combined with data from recent spectroscopic surveys, have shown that the halo and the thick disc carry the signatures of at least one merger of the Milky Way with a $\sim 10^{9-10}$\,\msun\ galaxy about 10\,Gyrs ago, and that these Galactic components largely constitute the debris from this collision \citep[see][for a brief recent review]{Wyse_2019}. \textit{Gaia} revealed that halo stars, both main-sequence and giants, are divided in the HRD into two parallel sequences, a ``blue'' low-metallicity sequence and a ``red'' higher-metallicity locus. \citet{Belokurov_2018}, \citet{Haywood_2018}, \citet{Gallart_2019}, and others, showed that the blue halo is largely composed of stars that were accreted from merged galaxies. Much, or perhaps all, of the red halo, in turn, is composed of thick-disc stars that were heated by the encounter \citep{DiMatteo_2019, Amarante_2020}, i.e. the red halo and the thick disc have essentially the same origin, with the thick disc itself probably being an ancient Milky Way pre-merger structure that was heated and thickened by the merger. A number of kinematically distinct retrograde halo structures have been tentatively identified as the stellar debris of the merged galaxy or galaxies themselves---named Gaia-Sausage \citep{Belokurov_2018}, Gaia-Enceladus \citep{Helmi_2018}, Gaia-Sequoia \citep{Myeong_2019}, and others. \citet{Feuillet_2020} showed that Gaia-Enceladus stars are associated mainly with the blue halo sequence. In terms of \af, \citet{Helmi_2018} and \citet{Mackereth_2019} showed that Gaia-Enceladus stars, in the \af\--[Fe/H] plane, are distributed differently from other halo and thick-disc stars, particularly having a larger spread toward low \af\ at higher [Fe/H]. This has been interpreted as the chemical signature of the accreted galaxy, with its distinct star-formation history.

An alternative interpretation of this stellar population posits an encounter between the Milky Way and the Andromeda galaxy some 10\,Gyr ago, which is responsible for the creation of the thick disc, the warp, and the bulge of the Milky Way, as well as its satellite galaxies \citep{Zhao_2013, Banik_2018, Bilek_2018}. The high \af\ of the thick disc stars is attributed to the time of the encounter, which took place before a substantial contribution to the iron budget of the Galaxy was made by SNe~Ia. In this scenario, the stellar halo might have formed \textit{in-situ} from a large population of embedded clusters, of which only the most massive ones have survived as present-day globular clusters \citep{Kroupa_2002, Baumgardt_2008}.

Apart from early-type galaxies, and in halo and thick-disc stars in the Milky Way, high \af\ has also been measured in some regions of M31---in its inner halo, its giant stellar streams, and its outer disc \citep{Escala_2020}. High-\af\ stars thus seem to appear in an array of galactic environments. 
The similarities between the stellar populations in massive elliptical galaxies and in the Milky Way's halo, in terms of high \af\ and SN Ia rate-dependent observables, raise the question of whether the IMF of halo stars (or their subset that has been possibly accreted from another galaxy) is also peculiar, and perhaps bottom-heavy, as in elliptical galaxies. 

In this paper, we use \textit{Gaia} DR2 to show that, indeed, the Milky Way's likely-accreted blue stellar halo has a bottom-heavy IMF. We also find it is distinct from the IMFs of the red halo and the thick disc---while the similarity of the IMFs of the latter two is in line with the likely common origin of these two components. The thin disc, in turn, has an IMF that is slightly different from the thick disc and the red halo. \textit{Gaia}, for the first time, permits the analysis of large (thousands of stars) and complete stellar samples selected according to kinematic component and metallicity, ideal for IMF determination. The IMF range below $\sim 1$\,\msun, with which we are concerned, is particularly straightforward to probe, as stars in that mass range are still in the main-sequence stage of their evolution, and therefore the current mass function and the IMF are one and the same (i.e. no accounting is needed for stars that have evolved post-main-sequence, which would necessitate assumptions about star-formation history). Thus, except for the need to deal with some observational effects (e.g. completeness, extinction, unresolved binaries), a count of stars as a function of their masses gives an almost direct measurement of the IMF.

\citet{Sollima_2019} has recently used \textit{Gaia} DR2 to measure the IMF of local stars, using a sample within 50\,pc for the IMF below 1\,\msun. His analysis however, does not separate stars according to Galactic kinematic and metallicity components, as we do here. Furthermore, he adopts a forward-modelling approach that differs from ours, and employs different procedures for dealing with some aspects of the problem, e.g. binarity. Our treatment thus also constitutes an independent analysis and a test of his results (a comparison is made in \S\ref{sec:PreviousWork}).

\section{Sample selection}
\label{sec:sample}

A primary concern in any IMF measurement is sample completeness. \citet{Sollima_2019} has demonstrated, by cross-matching stars from the \textit{Gaia} DR2 catalogue against the $3\pi$ Pan-STARRS \citep[PS1;][]{Flewelling_2018} and the Two-Micron All-Sky Survey \citep[2MASS;][]{Skrutskie_2006} catalogues, that \textit{Gaia} DR2 is $\gtrsim 90$~per cent complete in the magnitude range $7.5<G<18$, with completeness falling sharply at fainter magnitudes. We adopt $0.2$\,\msun\ as the approximate low-mass limit of stars to which we will investigate the IMF. This mass corresponds to an absolute magnitude of about 12 for main-sequence, Solar-metallicity stars, and $\sim 10-11$ for lower metallicities (see \S\ref{sec:mass}, below). These absolute magnitudes, and the $G<18$ mag completeness limit, together dictate samples with maximum distances of $\sim 100$\,pc for high-metallicity populations, and $\sim 250$\,pc for low-metallicity populations.

We therefore begin by retrieving all the \textit{Gaia} DR2 sources within 250\,pc, i.e. with DR2 parallax parameter satisfying
\begin{equation}
    \texttt{parallax} \geq 4,
\end{equation}
and that also satisfy the following astrometric and photometric accuracy criteria, to ensure a reliable positioning on the \textit{Gaia} Hertzsprung-Russell diagram (HRD):
\begin{align}
\label{eq:parallax_error}\texttt{parallax\_over\_error} & > 10\\
\texttt{phot\_g\_mean\_flux\_over\_error} & > 10 \\
\texttt{phot\_bp\_mean\_flux\_over\_error} & > 10 \\
\label{eq:phot_rp_error}\texttt{phot\_rp\_mean\_flux\_over\_error} & > 10.
\end{align}
These criteria limit the allowed relative error in the measured parallax, and in the $G$, $G_\textrm{BP}$, and $G_\textrm{RP}$ photometry.

While the $G$-band flux is measured by profile-fitting a narrow image, the flux in the $G_\textrm{BP}$ and $G_\textrm{RP}$ bands represents the total flux in an extended field. Thus, the \textit{Gaia} colour information is more vulnerable to contamination from nearby sources. In order to avoid sources with unreliable colour information, we limit the total $G_\textrm{BP}$ and $G_\textrm{RP}$ excess, compared to the $G$ band \citep{Evans_2018}:
\begin{align}
\texttt{phot\_bp\_rp\_excess\_factor} &> 1.0 + 0.015 \left(G_\textrm{BP}-G_\textrm{RP}\right)^2\\
\texttt{phot\_bp\_rp\_excess\_factor} &< 1.3 + 0.06 \left(G_\textrm{BP}-G_\textrm{RP}\right)^2.
\end{align}
We limit contamination by stars with flawed astrometric distances by requiring
\begin{equation}
    \texttt{ruwe} \leq 1.4,
\end{equation}
where \texttt{ruwe} is the re-normalized unit weight error, which is a measure of the astrometric goodness-of-fit statistic, re-normalized to eliminate magnitude and colour dependence \citep{Lindegren_2018, Lindegren_2018b}.
See Appendix~\ref{sec:GaiaQuery} for the full \textit{Gaia} ADQL query, resulting in a total of $2\,622\,304$ \textit{Gaia} sources.

Following \citet{Sollima_2019}, we correct the sample's photometry for extinction using a relation for the reddening,
\begin{equation}
    E \left(B-V\right) = \frac{0.03}{\sin b} \left[ \erf\left( \frac{\varpi^{-1} \sin b + z_\odot}{\sqrt{2} \sigma_\textrm{dust}} \right) - \erf \left( \frac{z_\odot}{\sqrt{2} \sigma_\textrm{dust}} \right) \right],
\end{equation}
where  $\varpi$ is the parallax, $b$ is the Galactic latitude, $z_\odot=1.4$\,pc is the Sun's height above the Galactic plane, and $\sigma_\textrm{dust}=150$\,pc.
The reddening in the \textit{Gaia} bands, $E\left(G_\textrm{BP}-G_\textrm{RP}\right)=\left(R_\textrm{BP} - R_\textrm{RP} \right)E \left(B-V\right)$, and the $G$-band extinction, $A_G=R_G E \left(B-V\right)$, are evaluated with the extinction coefficients $R_\textrm{BP}=3.374$, $R_\textrm{RP}=2.035$, and $R_G=2.740$ \citep[table 2]{Casagrande_2018}. Although this formulation for dust extinction is simple and approximate, at distances $< 250$\,pc we do not expect extreme extinction effects. Furthermore, a dust extinction vector on the HRD moves stars to fainter and redder magnitudes, roughly parallel to main-sequence-star curves of constant mass and age but of increasing metallicity (see \S\ref{sec:mass}, below). We therefore also do not expect a significant effect of extinction on the mass that we assign to a star based on its HRD position, and on the resulting IMF. Nonetheless, to gauge the effect of this approximation on our final results, in \S\ref{sec:alternatives}, below, we re-measure the IMF without any extinction correction, and also in various subsamples having different Galactic latitude and longitude limits. We find that, as expected, the extinction correction has a minor effect on our results.

From the apparent magnitude number distributions of the stars in the sample, we find that incompleteness, evident as a sharp cutoff in star counts, sets in at $G\gtrsim 18$, $G_\textrm{BP} \gtrsim 19.5$, and $G_\textrm{RP}\gtrsim 16.5$. For each star in our samples we calculate the absolute magnitude in the $G$ band,
\begin{equation}
    M_G = G + 5\times\log_{10}\left(\varpi\right)-10,
\end{equation}
and the transverse velocity,
\begin{equation}
    v_T = \frac{1}{\varpi} \sqrt{\mu_\textrm{RA}^2+\mu_\textrm{Dec}^2} 4.74\,\textrm{\kms},
\end{equation}
where $G$ is the $G$-band mean magnitude, $\varpi$ is the parallax in mas, and $\mu_\textrm{RA}$ and $\mu_\textrm{Dec}$ are the right ascension and declination proper motions in mas\,yr$^{-1}$, respectively.

Following \citet{Gaia_Babusiaux_2018}, we divide our initial sample into three subsamples based on the transverse velocity (see Fig.~\ref{fig:HRD}): ``thin disc'' ($v_T<40$\,\kms), ``thick disc'' ($60<v_T<150$\,\kms), and ``halo'' ($v_T>200$\,\kms). The halo subsample is further split in two, based on the location in the HRD, separating the ``blue'' and ``red'' sequences defined by \citet{Gaia_Babusiaux_2018} and discussed in \S\ref{sec:intro} above. This separation is consistent with a metallicity selection of stars with \mh~$\lessgtr-0.6$. These ``base samples''  that we have defined are listed in Table~\ref{tab:BaseSamples} and illustrated in Fig.~\ref{fig:HRD}.

\begin{table}
    \caption{\textit{Gaia} base samples, based on kinematics and metallicity.}
    \label{tab:BaseSamples}
    \centering
    \begin{tabular}{l c c c c }
        \hline
        Base sample & $v_{T\textrm{,min}}$ & $v_{T\textrm{,max}}$ & $\mhf_\textrm{min}$ & $\mhf_\textrm{max}$\\
         & (\kms) & (\kms) & & \\
        \hline
        Thin disc & $-$ & $40$ & $-1.5$ & $0.7$\\
        Thick disc & $60$ & $150$ & $-1.5$ & $0.7$\\
        High-metallicity halo & $200$ & $-$ & $-0.6$ & $0.7$\\
        Low-metallicity halo & $200$ & $-$ & $-2$ & $-0.6$\\
        High-metallicity halo 100kms & $100$ & $-$ & $-0.6$ & $0.7$\\
        Low-metallicity halo 100kms & $100$ & $-$ & $-2$ & $-0.6$\\
        All stars & $-$ & $-$ & $-2$ & $0.7$\\
        \hline
    \end{tabular}
\end{table}

\section{Mass assignment and binary correction}
\label{sec:massbinary}

\subsection{Mass assignment}
\label{sec:mass}

To derive an IMF, we need, in principle, to assign a mass to every star that we have included in our samples, whether on the main sequence or near it. The position of a star on the HRD depends on its mass, age, and metallicity, with some degeneracy among the three parameters, and hence one generally cannot determine uniquely a stellar mass based solely on HRD position. However, once they have evolved onto the main-sequence, stars move on the HRD only slightly with age, and their positions are largely determined by their masses (which vary along the main sequence), and their metallicities (which vary diagonally to the main sequence), as seen in Fig.~\ref{fig:PARSEC}. The mass of main-sequence stars can therefore be estimated quite accurately from their HRD position, as we further show below \citep[a similar approach was used by][]{Kroupa_1993}. The code used to derive the stellar parameters is provided in the \textsc{stam} (``Stellar-Track-based Assignment of Mass'') \textsc{python} package on \textsc{github}\footnote{\href{https://github.com/naamach/stam}{https://github.com/naamach/stam}}.

\begin{figure}
    \includegraphics[width=\columnwidth]{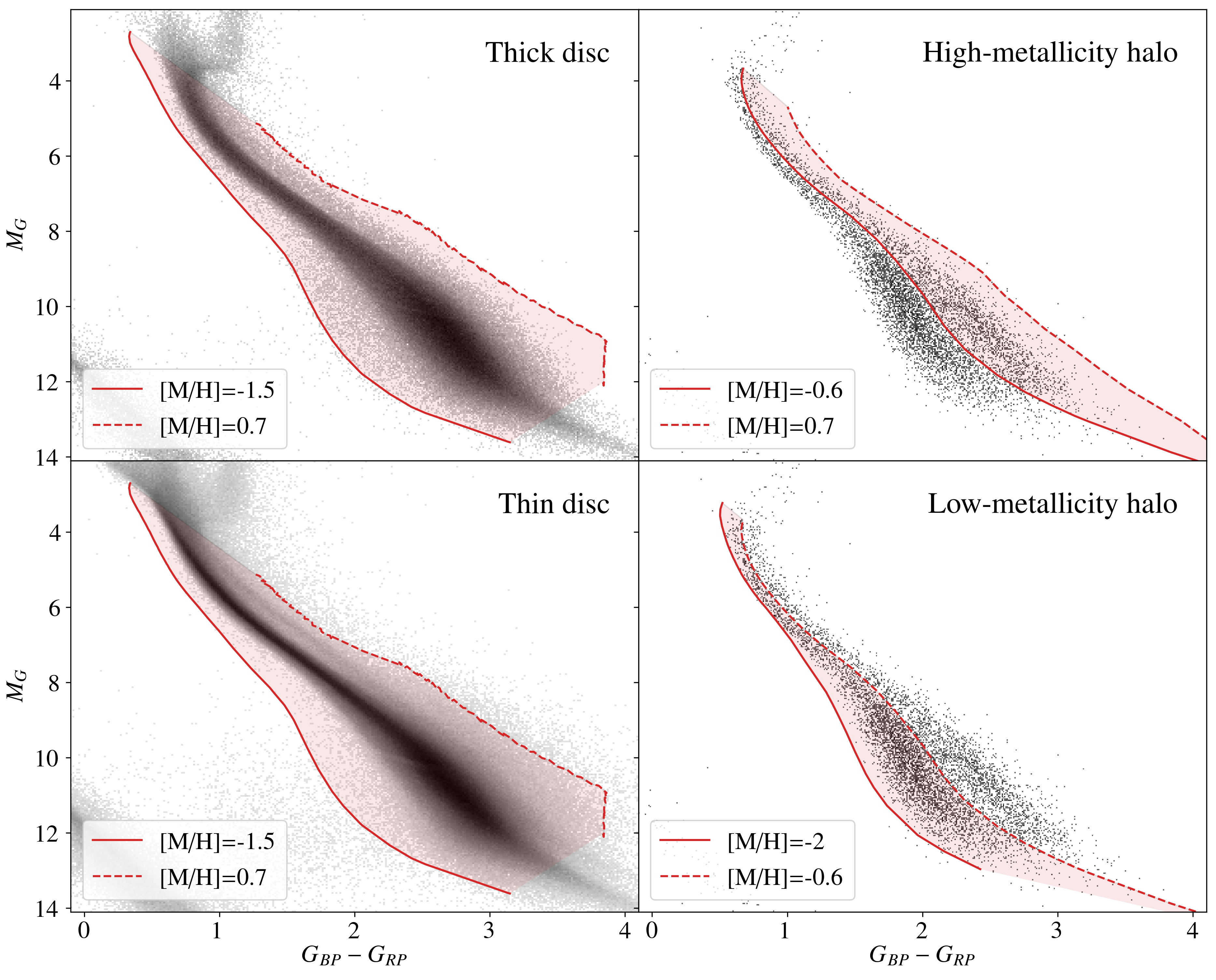}
    \caption{\textit{Gaia} HRD of the various kinematic subsamples, clockwise from bottom-left: Thin disc ($v_T<40$\,\kms), thick disc ($60<v_T<150$\,\kms), high-metallicity halo ($v_T>200$\,\kms), and low-metallicity halo ($v_T>200$\,\kms). The grayscale shows the \textit{Gaia} number density on the HRD of the sources in each kinematic subsample. The bounding PARSEC isochrones used to define the main sequence are plotted as red curves (see \S\ref{sec:mass} for details). The region used to compute each subsample's IMF is shaded red.}
    \label{fig:HRD}
\end{figure}

\begin{figure}
    \includegraphics[width=\columnwidth]{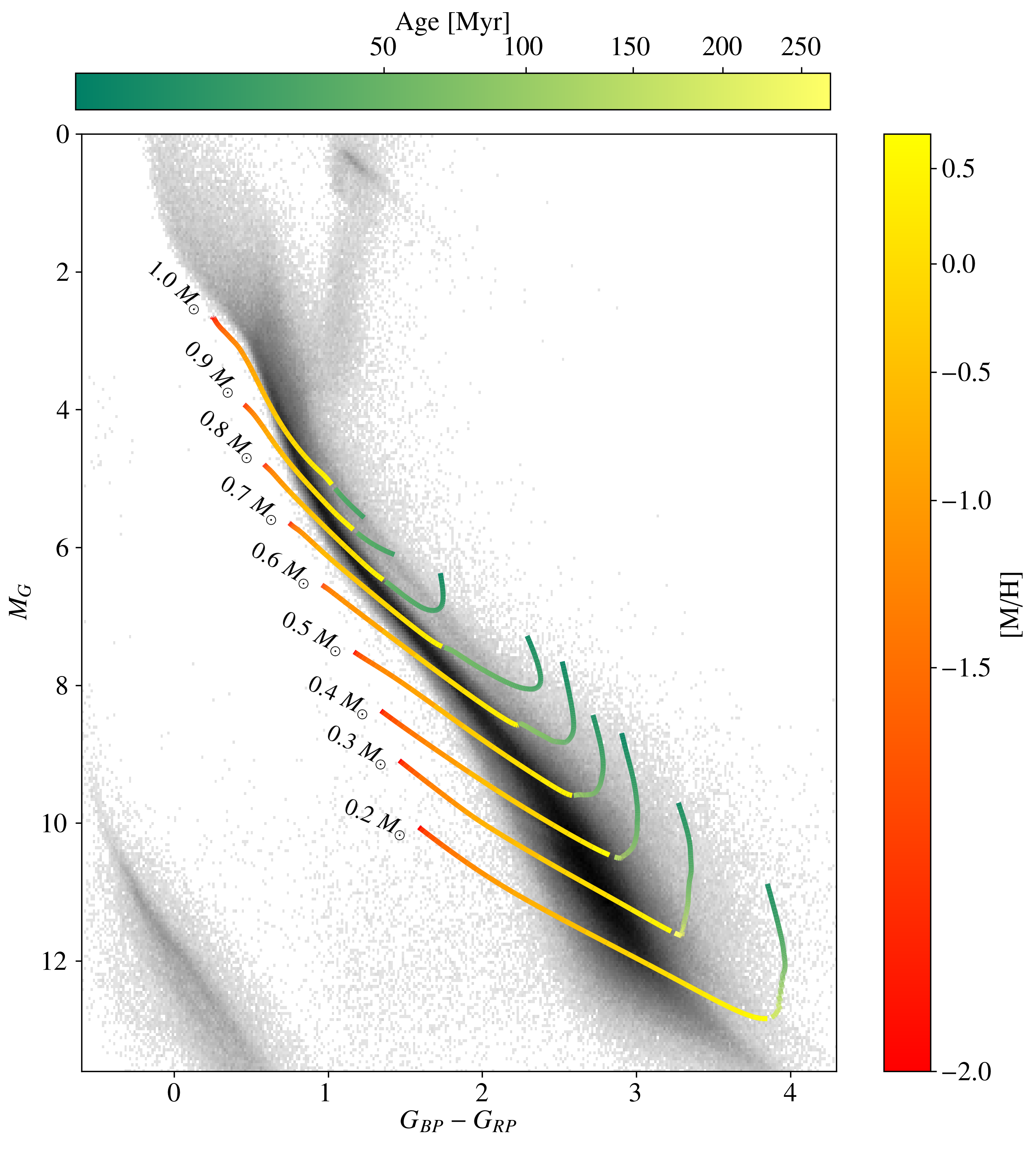}
    \caption{PARSEC isomass tracks used to interpolate the stellar masses for the thin-disc and thick-disc samples. Each isomass track is made of two parts: a fixed-metallicity (\mh~$=0.7$) pre-main-sequence track varying with age, and a fixed-age (5\,Gyr, for the disc samples) main-sequence track varying with metallicity (see \S\ref{sec:mass} for details). The HRD density of stars in the extinction-corrected thin disc 250pc sample is shown in grayscale.}
    \label{fig:PARSEC}
\end{figure}

Some of the stars seen to lie above the main sequence on the \textit{Gaia} HRD are pre-main-sequence stars. For pre-main-sequence stars, location on the HRD depends mainly on age and mass---a pre-main-sequence star of a given mass evolves roughly vertically towards the main sequence over several dozen Myr. An approximately correct mass can therefore be assigned for pre-main-sequence stars as well. However, some of the sources above the main sequence are actually unresolved binaries, in which the combined flux of the two stars raises them in the HRD above the main sequence. Indeed, a binary locus, composed of equal-mass binaries, is easy to discern in Figs.~\ref{fig:HRD} and \ref{fig:PARSEC}, $\sim 0.7$\,mag above the main sequence. We treat these equal-mass binaries separately, as described in \S\ref{sec:twins}. Finally, some stars seemingly above the main sequence, as predicted by stellar evolution models, may be the result of model inadequacies, such as ``radius inflation'', the under-prediction by models of the observed radii of M~dwarfs. Although the models we will use here have been empirically re-calibrated to deal with radius inflation, some remaining systematics could conceivably still result in some stars being above the model main sequence on the HRD \citep{Morrell_2019}.

With the above considerations in mind, we have used the PARSEC evolutionary tracks \citep[PAdova and tRieste Stellar Evolution Code;][version 1.2S]{Bressan_2012, Chen_2014, Chen_2015, Tang_2014}\footnote{\href{http://stev.oapd.inaf.it/cgi-bin/cmd_3.3}{http://stev.oapd.inaf.it/cgi-bin/cmd\_3.3}} to assign Galactic component membership (Fig.~\ref{fig:HRD}) and to assign masses to the \textit{Gaia} sources in our samples. As a first step, we select the main-sequence and pre-main-sequence stars in each subsample, by taking only the stars that are located on the \textit{Gaia} HRD between chosen PARSEC isochrones (see Fig.~\ref{fig:HRD}). For the thin and thick disc subsamples, we use the 5\,Gyr, \mh~$=-1.5$, main-sequence isochrone, and the 10\,Myr ,\mh~$=0.7$, pre-main-sequence isochrone, as the sample boundaries. For the halo subsamples, we use 10\,Gyr main-sequence isochrones of different metallicities for the boundaries: \mh~$=-2$ and $-0.6$ for the ``blue'', low-metallicity, halo subsample, and \mh~$=-0.6$ and $0.7$ for the ``red'', high-metallicity, halo subsample.

For the main-sequence stars in the PARSEC models (whose position on the HRD, as noted, does not change much with age) we have compiled a grid of metallicity tracks---i.e., tracks of constant age and mass, but varying metallicity---for masses ranging from 0.15 to 1.05\,\msun, in 0.05\,\msun\ steps. For the thin and thick disc subsamples we used a fixed age of 5\,Gyr, while for the halo subsamples we used a fixed age of 10\,Gyr. Using pre-main-sequence stars in the models, we have also created tracks of fixed mass and metallicity (\mh~$=0.7$), with age growing from zero to the age of the zero-age main-sequence. The combined pre-main-sequence (with varying age) and main-sequence (with varying metallicity) ``isomass'' tracks (i.e. tracks of the same mass) were smoothed to remove numerical artefacts. Fig.~\ref{fig:PARSEC} shows an example of the isomass tracks used to assign masses to the thin-disc subsample.

\begin{figure}
    \includegraphics[width=\columnwidth]{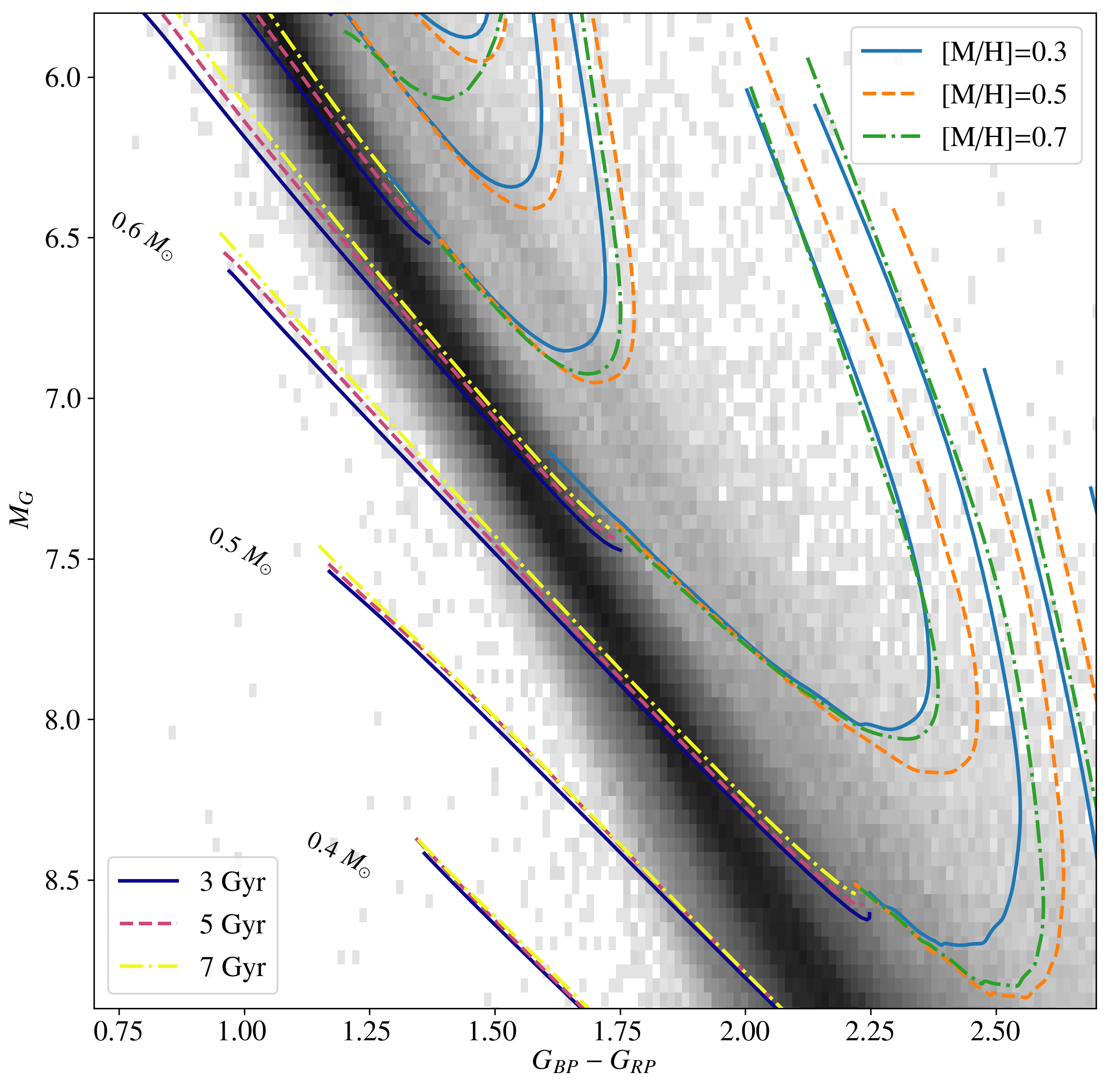}
    \caption{A zoomed version of Fig.~\ref{fig:PARSEC}, with varying age in the main-sequence part, and varying metallicity in the pre-main-sequence part. Within the ranges relevant for a given Galactic population (thin disc, in this figure), the changes in the isomass tracks when varying the model ages and metallicities are small.}
    \label{fig:IsomassZoom}
\end{figure}

To illustrate the robustness of our approach of mass assignment to variations in age in the case of main-sequence stars, and in metallicity in the case of pre-main-sequence stars, Fig.~\ref{fig:IsomassZoom} zooms in on some of the isomass tracks. The changes in the isomass tracks when varying the model ages and metallicities within ranges relevant for a given Galactic population, are small.

We use the grid of isomass curves to interpolate, between the curves, the masses and metallicities of each star in the subsample. The process is performed using \textsc{scipy}'s linear radial basis function (RBF) interpolation\footnote{\href{https://docs.scipy.org/doc/scipy/reference/generated/scipy.interpolate.Rbf.html}{https://docs.scipy.org/doc/scipy/reference/generated/scipy.interpolate.Rbf .html}}. In order to account for the \textit{Gaia} photometric and astrometric uncertainties, the location of each star on the HRD is drawn ten times from a two-dimensional normal distribution in the $\left(G_\textrm{bp}-G_\textrm{rp},~M_G\right)$ space, based on the star's photometric measurement errors.
The colour uncertainty is taken as
\begin{equation}
    \Delta (G_\textrm{bp}-G_\textrm{rp}) = 2.5 \log_{10}(e) \sqrt{\left(\frac{\Delta f_{G_\textrm{bp}}}{f_{G_\textrm{bp}}}\right)^2 + \left(\frac{\Delta f_{G_\textrm{rp}}}{f_{G_\textrm{rp}}}\right)^2},
\end{equation}
where $e$ is Euler's number, $f_i$ is the flux in the $i$-band; and the absolute magnitude uncertainty is calculated as
\begin{equation}
    \Delta (M_G) = 2.5 \log_{10}(e) \sqrt{\left(\frac{\Delta f_\textrm{G}}{f_\textrm{G}}\right)^2 + \left(2\frac{\Delta \varpi}{\varpi}\right)^2}.
\end{equation}
Each star is then assigned a mass, a metallicity, and uncertainties in these parameters, based on the mean and standard deviation of the ten realizations. We represent each star's contribution to the IMF with a unit-normalized Gaussian mass probability distribution having the measured mean and standard deviation of the mass. This error propagation procedure properly relates a model mass uncertainty to the photometric and astrometric errors for each star. We then calculate the IMF for each of our subsamples by summing all of the Gausssians into mass bins between 0.15 to 1.10\,\msun, in 0.05\,\msun\ steps.

In principle, the observed IMF calculated here is the convolution of the underlying intrinsic IMF with the observed mass errors. In order to reconstruct what is the intrinsic IMF, we would therefore need to fit the observed IMF with error-convolved intrinsic IMF functions. However, the intrinsic IMF is assumed to be a power law, and the mass errors are assumed to be symmetric Gaussians, approximately independent of the mass (see Fig.~\ref{fig:Errors}). Since the convolution of a power law with a relatively narrow Gaussian leaves the power law nearly unchanged, fitting the observed IMF directly should recover the intrinsic IMF power law. We have verified this by fitting the observed IMF with a power law convolved with a Gaussian (with a width equal to the median mass uncertainty), recovering the same power-law indices. The power laws that we fit to the observed IMF therefore do represent the intrinsic IMF.

Not at the focus of this paper, yet still of general interest beyond the IMF, our analysis also reveals the distributions of metallicity and transverse velocity for each of our complete local kinematic subsamples of stars, and the correlations between mass, metallicity, and velocity.  We present these in Appendix~\ref{sec:SampleProperties}.

\subsection{Binary correction}
\label{sec:binary}

\subsubsection{Equal-mass binaries}
\label{sec:twins}

\begin{figure}
    \includegraphics[width=\columnwidth]{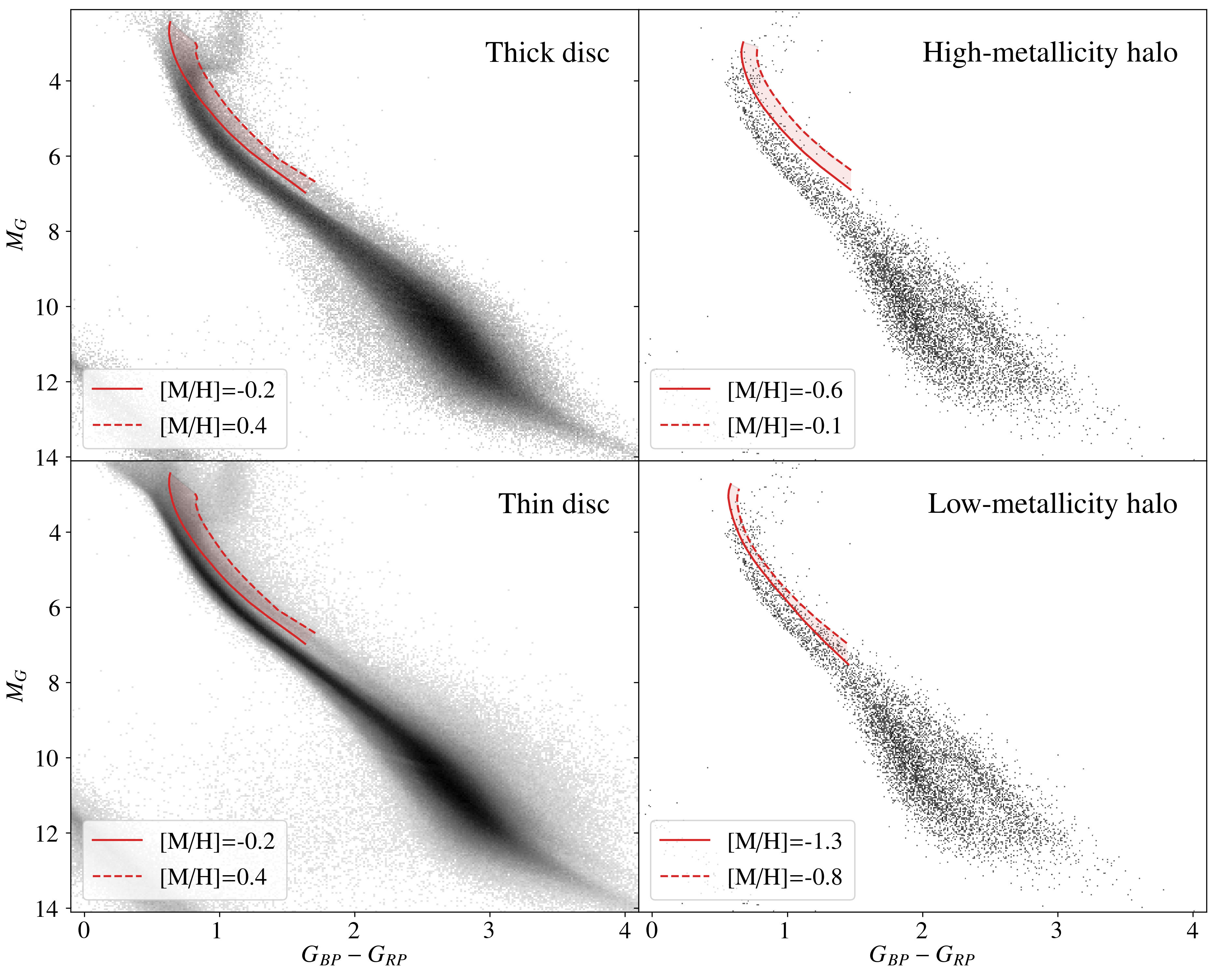}
    \caption{Same as Fig.~\ref{fig:HRD}, but with the equal-mass binary sequence shaded red. The bounding PARSEC isochrones used to define the equal-mass binary sequence are plotted as red curves (see \S\ref{sec:twins} for details). The equal-mass binary treatment is not applied to the low-metallicity halo, since its binary locus overlaps with the high-metallicity halo main sequence.}
    \label{fig:HRDbinary}
\end{figure}

Equal-mass binaries, that are located in the binary locus $\sim 0.7$\,mag above the main sequence (see Fig.~\ref{fig:HRDbinary}, and \citealt{ElBadry_2019}), are treated separately during the mass assignment procedure. First, we define the equal-mass binary sequence in each subsample as the region bounded by two main-sequence PARSEC isochrones, shifted upward to brighter magnitudes on the \textit{Gaia} HRD by a factor equivalent to a flux ratio of 1.9 and 2, respectively (i.e. 0.7 and 0.75\,mag, see Fig.~\ref{fig:HRDbinary}). For the thin and thick disc subsamples, we use the 5\,Gyr, \mh~$=-0.2$ and \mh~$=0.4$, main-sequence isochrones as the sample boundaries. For the halo subsamples, we use 10\,Gyr main-sequence isochrones of different metallicities for the boundaries: \mh~$=-1.3$ and $-0.8$ for the low-metallicity halo subsample, and \mh~$=-0.6$ and $-0.1$ for the high-metallicity halo subsample.

To avoid confusing main-sequence single stars with binaries, we truncate the equal-mass binary sequence region at $G_\textrm{BP} - G_\textrm{RP}=1.8$ for the disc subsamples, and 1.5 for the halo subsamples, where the equal-mass binary sequence begins to  overlap with the fainter part of the single-star main sequence. This is equivalent to a low-mass limit of about 0.6\,\msun\ for the disc subsamples, 0.55\,\msun\ for the high-metallicity halo subsample, and 0.45\,\msun\ for the low-metallicity halo subsample. We then check, for each \textit{Gaia} source during the mass and metallicity assignment process, which of its ten realizations fall inside the equal-mass binary region. If any, we draw an additional ten realizations, and the star is then assigned a ``binary probability'' based on the number of realizations (out of 20) that fell inside the binary region. The mean and standard deviations of the realizations that fall inside and outside of the binary region are calculated separately, where the $M_G$ magnitudes of the binary realizations are shifted to fainter magnitudes by $\sim 0.75$\,mag (assuming an equal-mass binary) before assigning the mass and metallicity. During the IMF construction, these stars contribute three times to the IMF, weighted by the computed binary probability: once as a single star, and twice as the two binary components.

Since in the low-metallicity halo subsample the equal-mass binary sequence overlaps with the high-metallicity-halo main sequence (see Fig.~\ref{fig:HRDbinary}), we do not apply this equal-mass binary treatment to the low-metallicity halo. Instead, we apply the full binary correction factor described below to this subsample. The number of stars in the halo subsamples is relatively small, and only a few stars are located in the equal-mass binary sequence of the high-metallicity halo, so we expect the contamination by low-metallicity halo equal-mass binaries of the high-metallicity halo to be negligible. This expectation is supported by the observed transverse velocity distribution of the two halo subsamples; if the high-metallicity halo subsample was significantly contaminated by equal-mass binaries from the low-metallicity halo, we would have seen their contribution to the transverse velocity distribution of the high-metallicity halo. However, as seen in Fig.~\ref{fig:Velocity}, any such contamination cannot be greater than a few stars (out of the 2712 stars in the 250\,pc extinction-corrected high-metallicity halo subsample). The deviation of the assigned mass of the equal-mass binaries from their true mass is typically $\lesssim 0.1$\,\msun\ in the low-metallicity halo and in the lower mass bins of the high-metallicity halo and disc subsamples (where they overlap with the fainter end of the main sequence). Furthermore the binary fraction is known to decrease with decreasing primary mass (see below), and only a fraction of binaries have mass ratios close to 1. We therefore expect that only of order a few percent of the stars in the regions of the HRD that do not receive the equal-mass treatment are being assigned masses that are systematically high by $\lesssim 0.1$\,\msun\, and that this should have a negligible effect on the derived IMF (see  Fig.~\ref{fig:IMFconstruction}, for a demonstration).

\subsubsection{Contribution by unseen companions}
\label{sec:secondaries}

The IMF, at this point, still needs to be corrected for the effect of non-equal-mass binaries (as well as equal-mass binaries of masses lower than the low-mass limit of the treated binary locus), as each unresolved binary has been counted in the IMF as a single star having the approximate mass of the photometric (and mass) primary star in the system \citep[see][for the first implementation of a binary-corrected IMF]{Kroupa_1993}. The secondary star in a binary is either lost in the glare of the more-massive star and thus has little effect on the estimated mass of the singly counted primary or, in nearly-equal-mass binaries, the summed light is counted as one star, with a slightly overestimated mass ($\lesssim 0.1$\,\msun). The angular separation at which binary pairs are typically resolved by \textit{Gaia} \textit{and} have full colour information, is 2.3\,arcsec, see \citet[fig.~9]{Arenou_2018}. The maximal unresolved projected separation, for an angular resolution of 2.3\,arcsec, is 115, 230, 460, 575\,au at distances of 50, 100, 200, and 250\,pc, respectively.

In order to account for the loss of the undetected secondary stars in unresolved binaries, we apply a correction factor to the IMF at mass-bin $m$, 
\begin{equation}\label{eq:binary}
    C_\textrm{bin} \left(m \right) = 1+\int_{q=m/M_\odot}^1 \frac{\dv N}{\dv m}\left(\frac{m}{q}\right) f_\textrm{bin} \left(\frac{m}{q}\right) P \left(q\right) \dv q ~/~ \frac{\dv N}{\dv m}\left(m\right) ,
\end{equation}
where $\frac{\dv N}{\dv m}\left(m\right)$ is the measured, pre-binary-corrected mass function, $f_\textrm{bin} \left(m\right)$ is the binary fraction as a function of mass (defined as the fraction of all systems---single-star systems and binary systems---that are in fact binary systems that are unresolved by \textit{Gaia}), and $P \left(q\right)$ is the distribution of the secondary-to-primary mass-ratio $q$ in binaries. For the binary fraction, at first considering all possible binary separations, we interpolate from fig.~12 of \citet[section 5.3.2]{Raghavan_2010}, between the measured binary fractions 0.35, 0.41, 0.50, for stars of masses 0.3, 0.8, 1.2\,\msun, respectively. Since binaries with large-enough separations do get resolved by \textit{Gaia} and their components get counted as individual sources, we further estimate the fraction of binaries that are unresolved in our \textit{Gaia} samples, based on fig.~13 of \citet{Raghavan_2010}, who represent the binary orbital period distribution as a log normal in $\log(P/1\textrm{d})$, with a mean of $5.03$ and a standard deviation of $2.28$. The unresolved binary fraction $f_\textit{bin}\left(m\right)$ is then obtained by multiplying the total binary fraction in each mass bin by the unresolved separation percentile at the maximum distance of the subsample (see above), assuming separation and orbital period are related through Kepler's Law.
For the mass-ratio distribution, $P \left(q\right)$, we use the recent \textit{Gaia}-based measurements of \citet{ElBadry_2019} for resolved binaries. 
Given the wide log-normal separation distribution, the typical separation in all of our subsamples is $\lesssim 250$\,au. We therefore interpolate $P\left(q\right)$ from table~G1 of \citet{ElBadry_2019} for the separation range $50-350$\,au, which applies to much of the separation range of the binaries in our samples. In this $P\left(q\right)$ distribution, primaries less massive than 0.6\,\msun\ have a rather flat mass-ratio distribution, while more massive primaries ($0.6<m/M_\odot<1.2$) have a $P\left(q\right)$ with a peak near a mass-ratio of $0.5$. In addition, $P\left(q\right)$ for primaries of all masses includes a ``twin peak'' in the distribution---an excess of equal-mass companions, as described by \citet{ElBadry_2019}. We exclude this ``twin peak'' from the binary correction when we apply the special equal-mass binary treatment described above.
We note that our binary correction procedure consistently accounts for the contributions of the binary companion stars to the mass function, which has not always been the case in previous work \citep[e.g. in the high-mass regime of][]{Sollima_2019}.

The binary correction factor, as calculated above, is applied to the observed IMF, to obtain our final IMF for a given subsample. Since unresolved binaries tend to hide low-mass secondaries in the glare of the higher-mass primaries, the binary correction $C_\textrm{bin}$ is a decreasing function of mass \citep[see also][]{Kroupa_2019}. For example, for the thin disc 250\,pc subsample, $C_\textrm{bin}\left(m\right)$ decreases from $\sim 1.3$ at the low-mass end of the IMF, to only $\sim 1.01$ at $\sim$ 1\msun.

\begin{figure}
    \includegraphics[width=\columnwidth]{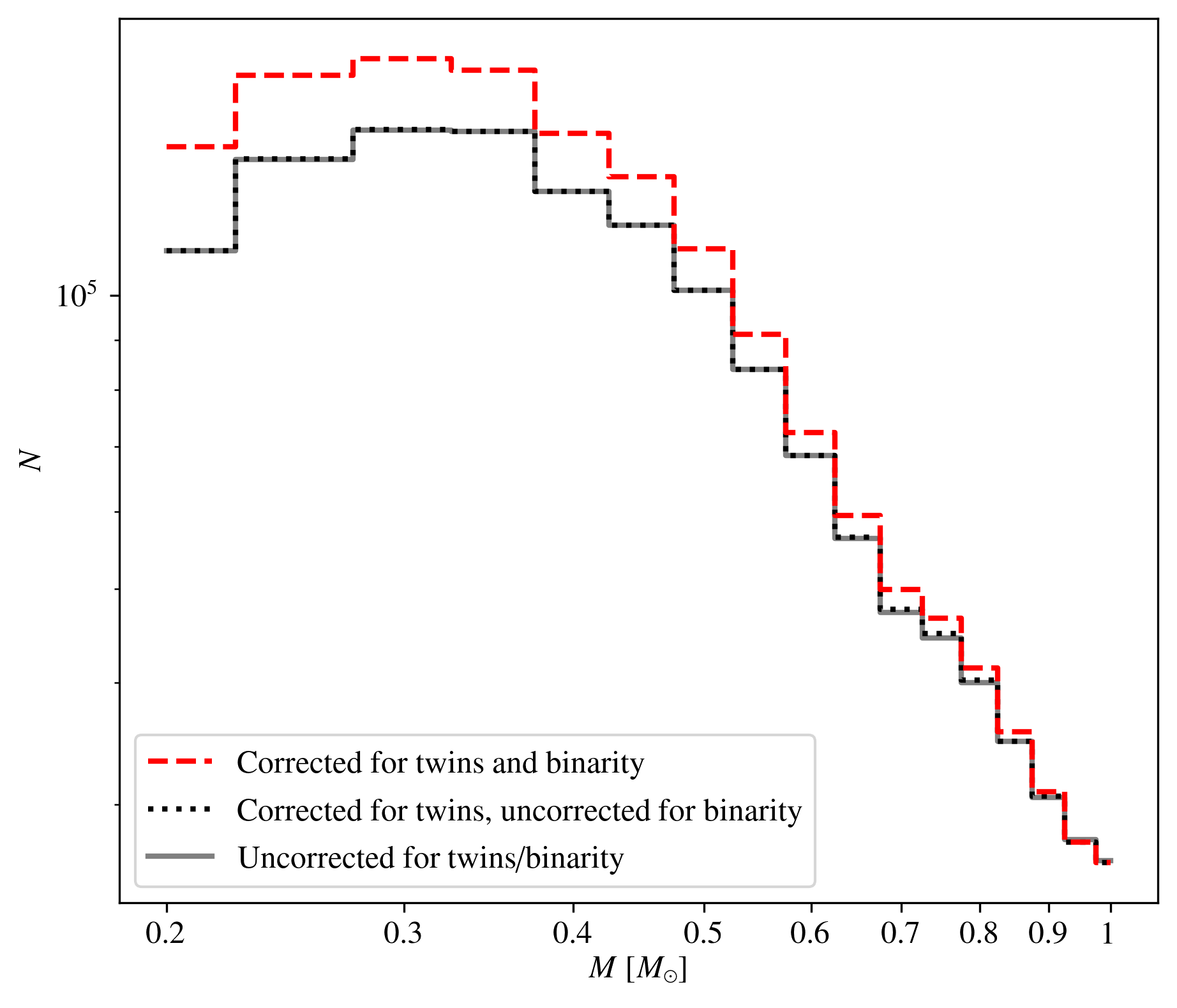}
    \caption{Example of IMF derivation, for the extinction-corrected 250\,pc thin disc subsample: IMF, constructed by summing the Gaussian mass probability distributions of all stars, uncorrected for equal-mass binaries (``twins'') or binarity (solid grey); corrected for twins but not binarity (dotted black); and after applying the binary correction, $C_\textrm{bin}$ (dashed red).}
    \label{fig:IMFconstruction}
\end{figure}

\begin{figure}
    \includegraphics[width=\columnwidth]{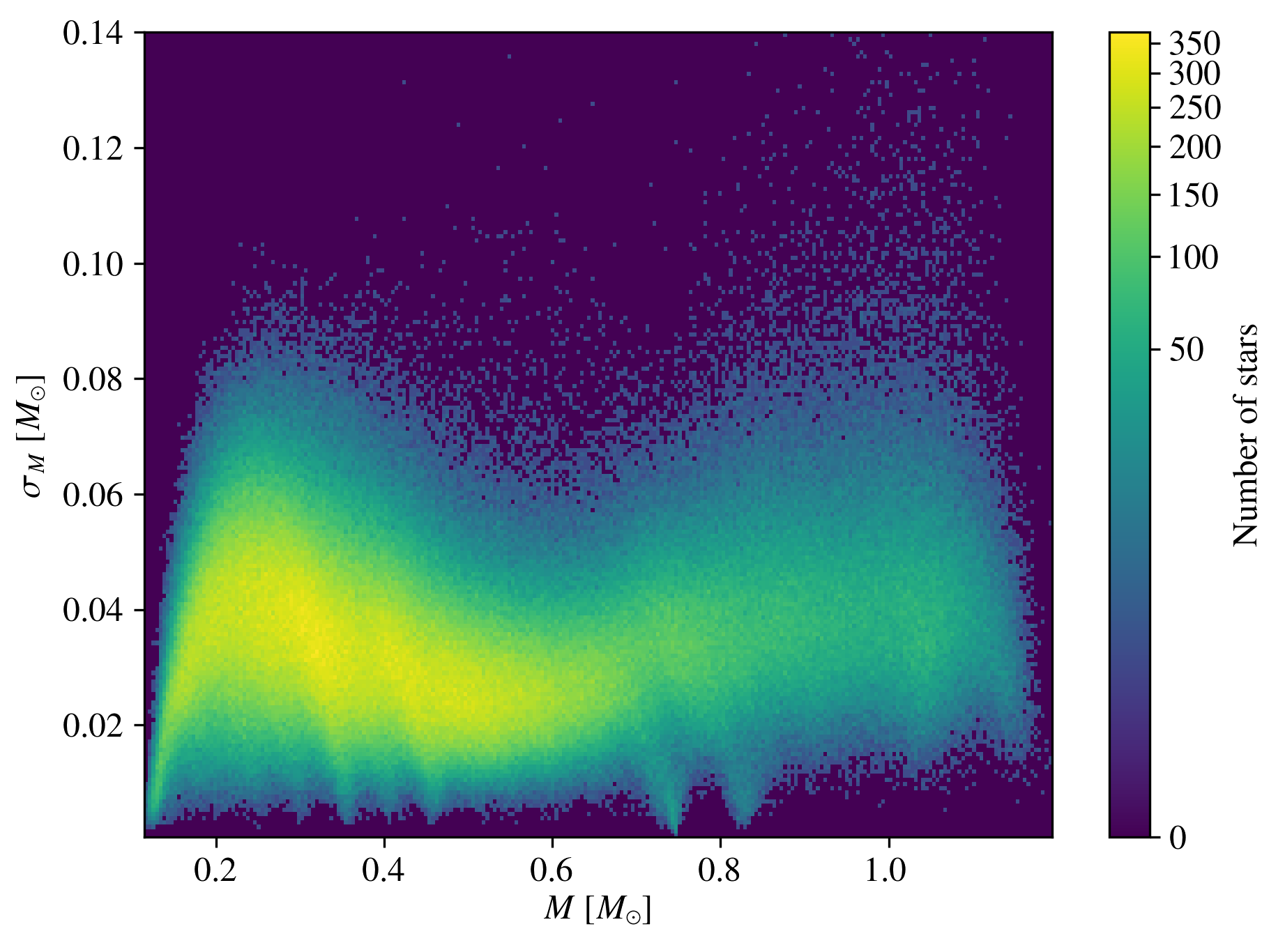}
    \caption{Two-dimensional histogram of the mass uncertainty vs. assigned stellar mass of the extinction- and twin-corrected 250\,pc thin disc subsample. The barely discernible arcs are artifacts of the mass-assignment interpolation process.}
    \label{fig:Errors}
\end{figure}

To illustrate our procedure for mass assignment and IMF measurement, we show in Fig.~\ref{fig:IMFconstruction}, for one of the subsamples, the IMF from the sum of the many individual Gaussian mass probability distributions representing all of the stars, the IMF corrected for equal-mass binaries, and the IMF after application of the binary correction $C_\textrm{bin}$. There is a negligible difference between the IMFs corrected and uncorrected for equal-mass binaries. The distribution of mass uncertainties around the estimated mass means is shown in Fig.~\ref{fig:Errors}.

\subsection{Testing result robustness using alternative subsamples}
\label{sec:alternatives}

To get a sense of the unknown systematics that affect our IMF estimates and to thus estimate the accuracy to which we can determine the IMF, we have defined a number of alternative subsamples, and repeated for each of them the analysis described above. The alternative subsamples have: different volumes (outer radii between 50 and 250\,pc); different Galactic directions (high latitude, low latitude, Galactic centre and anti-centre hemispheres); different transverse velocity thresholds for the halo subsample (100\,\kms\ instead of 200\,\kms); and subsamples with and without extinction, binarity, and equal-mass binaries correction. We have also tested the effect of applying stricter quality cuts, with minimal relative errors of 5~per cent instead of 10~per cent for the \textit{Gaia} photometric and astrometric measurements (Eqs. \ref{eq:parallax_error}$-$\ref{eq:phot_rp_error}).
The effect of the Lutz-Kelker bias \citep{Lutz_1973, Luri_2018}, causing the mean observed parallax to be systematically larger than its true value, is tested by comparing the results of the subsamples with different volumes, as well as those with stricter quality cuts.
Table~\ref{tab:subsamples} summarizes the various subsamples that we have explored.

\begin{table*}
    \caption{\textit{Gaia} subsamples and IMF power-law indices. The ``Extinction'' (``Binarity'') column denotes whether the subsample was corrected for extinction (binarity). The ``Twins'' column denotes whether the equal-mass binary sequence was treated separately during the mass assignment procedure (see \S\ref{sec:twins} for details). The ``Q?'' column denotes whether stricter \textit{Gaia} quality cuts were applied (see \S\ref{sec:alternatives} for details). $N_\textrm{stars}$ indicates the number of \textit{Gaia} sources included in each subsample. $\alpha_\textrm{low}$ ($\alpha_\textrm{high}$) is the IMF power-law index for masses lower (higher) than $\sim 0.5$\,\msun.}
    \label{tab:subsamples}
    \centering
    \begin{tabular}{l l c c c c c c c r l l}
        \hline
        ID & Base sample & $d$ & Extinction & Binarity & Twins & Q? & $b$ & $l$ & $N_\textrm{stars}$ & $\alpha_\textrm{low}$ & $\alpha_\textrm{high}$\\
           & & (pc) & & & & & (deg) & (deg) & & & \\
        \hline
        1 & Thin disc & 250 & $-$ & $-$ & $-$ & $-$ & all & all & $1470381$ & $1.10 \pm 0.01$ & $2.03 \pm 0.01$ \\
        2 & Thin disc & 250 & $+$ & $-$ & $-$ & $-$ & all & all & $1473916$ & $1.01 \pm 0.01$ & $2.00 \pm 0.01$ \\
        3 & Thin disc & 250 & $+$ & $-$ & $+$ & $-$ & all & all & $1473916$ & $1.00 \pm 0.01$ & $2.00 \pm 0.01$ \\
        4 & Thin disc & 250 & $+$ & $+$ & $-$ & $-$ & all & all & $1473916$ & $1.14 \pm 0.01$ & $2.10 \pm 0.01$ \\
        5 & Thin disc & 250 & $+$ & $+$ & $+$ & $-$ & all & all & $1473916$ & $1.13 \pm 0.01$ & $2.14 \pm 0.01$ \\
        6 & Thin disc & 250 & $-$ & $+$ & $+$ & $-$ & all & all & $1470381$ & $1.23 \pm 0.01$ & $2.17 \pm 0.01$ \\
        7 & Thin disc & 250 & $+$ & $+$ & $+$ & $-$ & $|b|\leq20$ & all & $619211$ & $1.14 \pm 0.02$ & $2.13 \pm 0.01$ \\
        8 & Thin disc & 250 & $+$ & $+$ & $+$ & $-$ & $|b|>20$ & all & $854705$ & $1.13 \pm 0.01$ & $2.15 \pm 0.01$ \\
        9 & Thin disc & 250 & $+$ & $+$ & $+$ & $-$ & all & $270 \leq l <90$ & $742969$ & $1.13 \pm 0.01$ & $2.16 \pm 0.01$ \\
        10 & Thin disc & 250 & $+$ & $+$ & $+$ & $-$ & all & $90\leq l<270$ & $730947$ & $1.14 \pm 0.01$ & $2.13 \pm 0.01$ \\
        11 & Thin disc & 100 & $-$ & $+$ & $+$ & $-$ & all & all & $119231$ & $1.37 \pm 0.04$ & $2.04 \pm 0.02$ \\
        12 & Thin disc & 100 & $+$ & $+$ & $+$ & $-$ & all & all & $119745$ & $1.34 \pm 0.04$ & $2.03 \pm 0.02$ \\
        13 & Thin disc & 100 & $+$ & $+$ & $+$ & $-$ & $|b|\leq20$ & all & $45302$ & $1.39 \pm 0.06$ & $2.04 \pm 0.04$ \\
        14 & Thin disc & 100 & $+$ & $+$ & $+$ & $-$ & $|b|>20$ & all & $74443$ & $1.30 \pm 0.05$ & $2.02 \pm 0.03$ \\
        15 & Thin disc & 100 & $+$ & $+$ & $+$ & $-$ & all & $270 \leq l <90$ & $60546$ & $1.34 \pm 0.05$ & $2.05 \pm 0.03$ \\
        16 & Thin disc & 100 & $+$ & $+$ & $+$ & $-$ & all & $90\leq l<270$ & $59199$ & $1.33 \pm 0.05$ & $2.01 \pm 0.03$ \\
        17 & Thin disc & 100 & $+$ & $+$ & $+$ & $+$ & all & all & $118079$ & $1.33 \pm 0.04$ & $2.03 \pm 0.02$ \\
        18 & Thick disc & 250 & $-$ & $+$ & $+$ & $-$ & all & all & $425944$ & $0.99 \pm 0.01$ & $2.51 \pm 0.01$ \\
        19 & Thick disc & 250 & $+$ & $+$ & $-$ & $-$ & all & all & $426713$ & $0.91 \pm 0.01$ & $2.46 \pm 0.01$ \\
        20 & Thick disc & 250 & $+$ & $+$ & $+$ & $-$ & all & all & $426713$ & $0.91 \pm 0.01$ & $2.51 \pm 0.01$ \\
        21 & Thick disc & 250 & $+$ & $+$ & $+$ & $-$ & $|b|\leq20$ & all & $125204$ & $0.94 \pm 0.02$ & $2.48 \pm 0.03$ \\
        22 & Thick disc & 250 & $+$ & $+$ & $+$ & $-$ & $|b|>20$ & all & $301509$ & $0.90 \pm 0.01$ & $2.52 \pm 0.02$ \\
        23 & Thick disc & 250 & $+$ & $+$ & $+$ & $-$ & all & $270 \leq l <90$ & $225648$ & $0.91 \pm 0.02$ & $2.50 \pm 0.02$ \\
        24 & Thick disc & 250 & $+$ & $+$ & $+$ & $-$ & all & $90\leq l<270$ & $201065$ & $0.91 \pm 0.02$ & $2.53 \pm 0.02$ \\
        25 & Thick disc & 100 & $-$ & $+$ & $+$ & $-$ & all & all & $30969$ & $1.17 \pm 0.05$ & $2.36 \pm 0.06$ \\
        26 & Thick disc & 100 & $+$ & $+$ & $+$ & $-$ & all & all & $31133$ & $1.14 \pm 0.05$ & $2.35 \pm 0.05$ \\
        27 & Thick disc & 100 & $+$ & $+$ & $+$ & $-$ & $|b|\leq20$ & all & $8988$ & $1.18 \pm 0.09$ & $2.35 \pm 0.10$ \\
        28 & Thick disc & 100 & $+$ & $+$ & $+$ & $-$ & $|b|>20$ & all & $22145$ & $1.12 \pm 0.05$ & $2.35 \pm 0.07$ \\
        29 & Thick disc & 100 & $+$ & $+$ & $+$ & $-$ & all & $270 \leq l <90$ & $16134$ & $1.17 \pm 0.06$ & $2.40 \pm 0.08$ \\
        30 & Thick disc & 100 & $+$ & $+$ & $+$ & $-$ & all & $90\leq l<270$ & $14999$ & $1.10 \pm 0.07$ & $2.29 \pm 0.08$ \\
        31 & High-metallicity halo & 250 & $-$ & $+$ & $+$ & $-$ & all & all & $2886$ & $1.18 \pm 0.15$ & $2.43 \pm 0.18$ \\
        32 & High-metallicity halo & 250 & $+$ & $+$ & $-$ & $-$ & all & all & $2712$ & $1.13 \pm 0.15$ & $2.66 \pm 0.19$ \\
        33 & High-metallicity halo & 250 & $+$ & $+$ & $+$ & $-$ & all & all & $2712$ & $1.07 \pm 0.15$ & $2.76 \pm 0.19$ \\
        34 & High-metallicity halo & 250 & $+$ & $+$ & $+$ & $-$ & $|b|\leq20$ & all & $692$ & $0.96 \pm 0.30$ & $2.48 \pm 0.35$ \\
        35 & High-metallicity halo & 250 & $+$ & $+$ & $+$ & $-$ & $|b|>20$ & all & $2020$ & $1.10 \pm 0.17$ & $2.86 \pm 0.22$ \\
        36 & High-metallicity halo & 250 & $+$ & $+$ & $+$ & $-$ & all & $270 \leq l <90$ & $1476$ & $0.84 \pm 0.20$ & $2.98 \pm 0.25$ \\
        37 & High-metallicity halo & 250 & $+$ & $+$ & $+$ & $-$ & all & $90\leq l<270$ & $1236$ & $1.33 \pm 0.23$ & $2.45 \pm 0.29$ \\
        38 & High-metallicity halo & 200 & $-$ & $+$ & $+$ & $-$ & all & all & $1543$ & $1.29 \pm 0.20$ & $2.61 \pm 0.26$ \\
        39 & High-metallicity halo & 200 & $+$ & $+$ & $+$ & $-$ & all & all & $1465$ & $1.17 \pm 0.21$ & $2.91 \pm 0.27$ \\
        40 & High-metallicity halo & 200 & $+$ & $+$ & $+$ & $-$ & $|b|\leq20$ & all & $355$ & $1.21 \pm 0.43$ & $2.59 \pm 0.53$ \\
        41 & High-metallicity halo & 200 & $+$ & $+$ & $+$ & $-$ & $|b|>20$ & all & $1110$ & $1.15 \pm 0.24$ & $3.01 \pm 0.31$ \\
        42 & High-metallicity halo & 200 & $+$ & $+$ & $+$ & $-$ & all & $270 \leq l <90$ & $789$ & $0.76 \pm 0.28$ & $3.36 \pm 0.36$ \\
        43 & High-metallicity halo & 200 & $+$ & $+$ & $+$ & $-$ & all & $90\leq l<270$ & $676$ & $1.60 \pm 0.32$ & $2.25 \pm 0.41$ \\
        44 & High-metallicity halo 100kms & 250 & $+$ & $+$ & $+$ & $-$ & all & all & $73423$ & $1.05 \pm 0.03$ & $2.52 \pm 0.03$ \\
        45 & Low-metallicity halo & 250 & $-$ & $+$ & $-$ & $-$ & all & all & $4293$ & \multicolumn{2}{c}{$1.90 \pm 0.04$} \\
        46 & Low-metallicity halo & 250 & $+$ & $+$ & $+$ & $-$ & all & all & $4473$ & \multicolumn{2}{c}{$1.84 \pm 0.04$} \\
        47 & Low-metallicity halo & 250 & $+$ & $+$ & $-$ & $-$ & all & all & $4473$ & \multicolumn{2}{c}{$1.82 \pm 0.04$} \\
        48 & Low-metallicity halo & 250 & $+$ & $+$ & $-$ & $-$ & all & all & $4473$ & $1.51 \pm 0.06$ & $2.97 \pm 0.21$ \\
        49 & Low-metallicity halo & 250 & $+$ & $+$ & $-$ & $-$ & $|b|\leq20$ & all & $1157$ & \multicolumn{2}{c}{$1.90 \pm 0.08$} \\
        50 & Low-metallicity halo & 250 & $+$ & $+$ & $-$ & $-$ & $|b|>20$ & all & $3316$ & \multicolumn{2}{c}{$1.79 \pm 0.05$} \\
        51 & Low-metallicity halo & 250 & $+$ & $+$ & $-$ & $-$ & all & $270 \leq l <90$ & $2321$ & \multicolumn{2}{c}{$1.82 \pm 0.06$} \\
        52 & Low-metallicity halo & 250 & $+$ & $+$ & $-$ & $-$ & all & $90\leq l<270$ & $2152$ & \multicolumn{2}{c}{$1.81 \pm 0.06$} \\
        53 & Low-metallicity halo & 200 & $-$ & $+$ & $-$ & $-$ & all & all & $2273$ & \multicolumn{2}{c}{$1.88 \pm 0.06$} \\
        54 & Low-metallicity halo & 200 & $+$ & $+$ & $-$ & $-$ & all & all & $2353$ & \multicolumn{2}{c}{$1.84 \pm 0.06$} \\
        55 & Low-metallicity halo & 200 & $+$ & $+$ & $-$ & $-$ & $|b|\leq20$ & all & $620$ & \multicolumn{2}{c}{$1.98 \pm 0.12$} \\
        56 & Low-metallicity halo & 200 & $+$ & $+$ & $-$ & $-$ & $|b|>20$ & all & $1733$ & \multicolumn{2}{c}{$1.79 \pm 0.06$} \\
        57 & Low-metallicity halo & 200 & $+$ & $+$ & $-$ & $-$ & all & $270 \leq l <90$ & $1260$ & \multicolumn{2}{c}{$1.90 \pm 0.08$} \\
        58 & Low-metallicity halo & 200 & $+$ & $+$ & $-$ & $-$ & all & $90\leq l<270$ & $1093$ & \multicolumn{2}{c}{$1.76 \pm 0.08$} \\
        59 & Low-metallicity halo 100kms & 250 & $+$ & $+$ & $-$ & $-$ & all & all & $9267$ & \multicolumn{2}{c}{$1.86 \pm 0.03$} \\
        60 & All & 50 & $+$ & $+$ & $+$ & $-$ & all & all & $22856$ & $1.31 \pm 0.08$ & $2.02 \pm 0.05$ \\
        61 & All & 50 & $+$ & $+$ & $-$ & $-$ & all & all & $22856$ & $1.23 \pm 0.08$ & $2.03 \pm 0.05$ \\
        \hline
    \end{tabular}
\end{table*}

\section{Results}
\label{sec:results}
\subsection{IMFs by Galactic component}

\begin{figure*}
    \includegraphics[width=\textwidth]{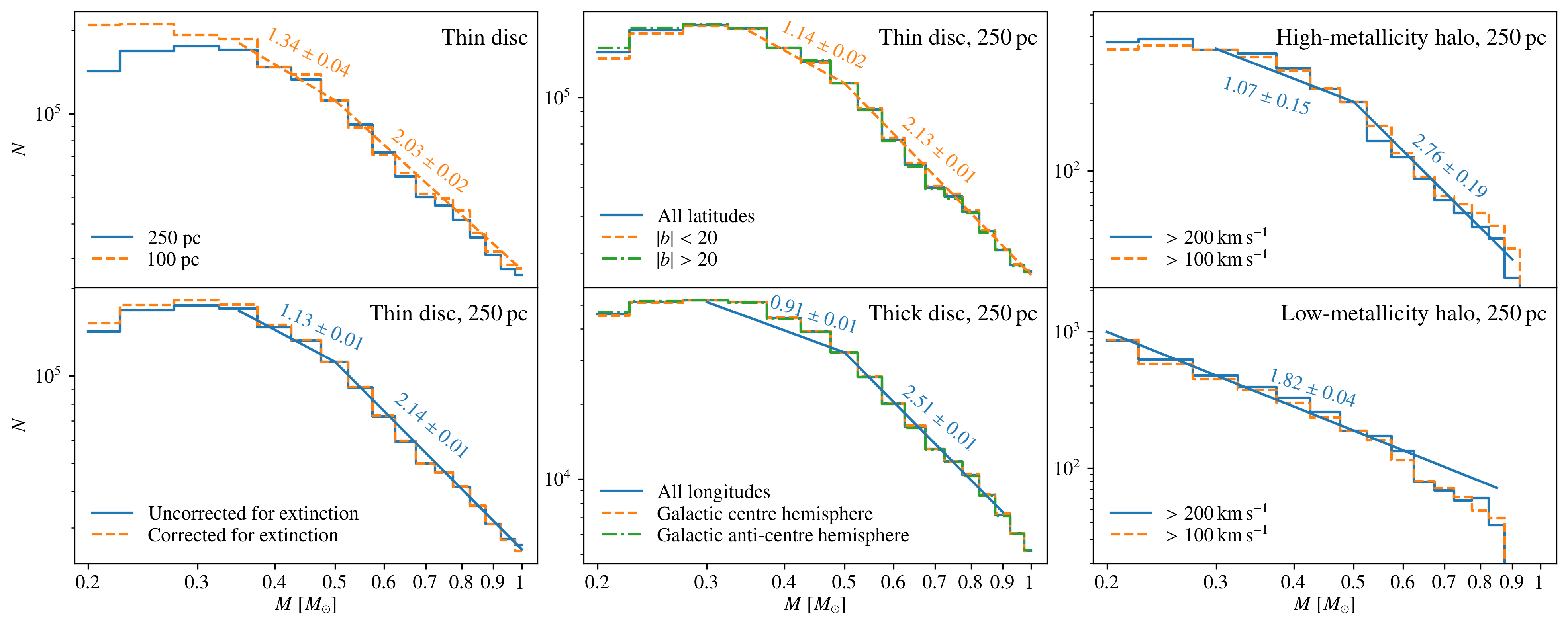}
    \caption{IMF of some individual subsamples. Top left: different sample volume comparison for the thin disc sample. Bottom left: the effect of extinction correction, for the thin disc 250\,pc sample. Top middle: high and low Galactic latitudes comparison, for the thin disc 250\,pc sample. Bottom middle: centre and anti-centre Galactic longitudes comparison, for the thick disc 250\,pc sample. Right: the effect of the transversal velocity threshold on the high (top) and low (bottom) metallicity halo 250\,pc samples. All displayed subsamples are corrected for extinction and binarity (including a separate mass assignment for the equal-mass binaries, except for the low-metallicity halo subsamples), unless noted otherwise. The power-law fit of one of the plotted subsamples in each subplot is represented by a line of the same style and colour. The fitted slope is indicated by the number above it.}
    \label{fig:IndividualIMF}
\end{figure*}

The IMFs that we have obtained for each of the four Galactic stellar component subsamples---thin disc, thick disc, high-metallicity halo, and low-metallicity halo---are shown in Fig.~\ref{fig:IndividualIMF}, along with the IMFs from some of the alternative subsamples we have used to estimate the systematic uncertainties in each component's IMF. 
Some general features, resulting from incompleteness effects, are visible in several of the IMFs. Stars of a given mass become fainter and redder with increasing metallicity, a result of their higher atmospheric opacities (see Fig.~\ref{fig:PARSEC}). The flattening toward the low-mass end in the IMFs of the high-metallicity Galactic components (i.e. all but the low-metallicity halo) is a result, at least partly, of this sample incompleteness to faint and red, low-mass, metal-rich, stars. This is evident, e.g., from a comparison of the IMFs based on the the 100\,pc and 250\,pc thin-disc samples in Fig.~\ref{fig:IndividualIMF}---in the more-distant 250\,pc sample, the incompleteness break naturally sets in at higher masses. Specifically, the apparent magnitude cutoffs of $G\gtrsim 18$, $G_\textrm{BP} \gtrsim 19.5$, and $G_\textrm{RP}\gtrsim 16.5$ mean that faint stars redder than $G_\textrm{BP} - G_\textrm{RP}\gtrsim 3$ will tend to be excluded from the sample, corresponding to masses smaller than $\sim 0.1$\,\msun\ at a distance of 100\,pc, and $\sim0.35$\,\msun\ at a distance of 250\,pc (see Fig.~\ref{fig:PARSEC}). This will deplete the $0.20-0.35$\,\msun\ bins of a large fraction of their main sequence stars. In addition, the HRD faintest and reddest part of the disc subsamples, partly contributing to the 0.2\,\msun\ pre-main-sequence region, is excluded based on the adopted definition of the main sequence region (see Fig.~\ref{fig:HRD}). For the halo components' IMF, a different kind of incompleteness appears, this time at the high-mass end. The main-sequence lifetime of a star decreases with metallicity. In the halo population, whose age is $\sim 10$\,Gyr, the stars with the lowest metallicities and with masses $\gtrsim 0.8$\,\msun\ have already evolved off the main sequence and are therefore absent from the sample. This is evident as a cutoff in the IMFs, at $\sim 0.8$\,\msun\ for the high-metallicity halo, and at $\sim 0.9$\,\msun\ for the low-metallicity halo. 

Fig.~\ref{fig:IndividualIMF} shows that, apart from the sample-volume-dependent low-mass cutoff, the IMF of each Galactic component is robust to changes in sample selection. As expected (see \S\ref{sec:sample}), the extinction correction has a minor effect on the IMF, and even a sample that is completely uncorrected for extinction differs from the corrected sample's IMF only at the lowest masses, already within the volume-incompleteness range. The IMFs change negligibly among samples limited to specific ranges of Galactic directions, whether low and high latitude, or Galactic centre and anti-centre longitudes. For the halo samples, the IMF likewise does not vary much if the threshold transverse velocity for halo membership is changed from 200\,\kms\ down to 100\,\kms. 

With the low-end and high-end IMF cutoffs in mind, we have fitted single or two-segment power laws to the IMF range that is within those turnover points and is unaffected by them. The fitted range is $0.35-1$\,\msun\ for the thin disc subsamples, $0.3-0.9$\,\msun\ for the thick disc and high-metallicity halo subsamples, and $0.2-0.85$\,\msun\ for the low-metallicity halo subsamples. Within those mass ranges, we have identified one (for single power laws) or two (for two-segment power laws) ranges in which the IMF appears as a straight line in the $\log \dv N/\dv m$ vs. $\log m$ plane. We fit each IMF for the power-law indices of the low-mass and high-mass segments, using $\chi^2$ minimization technique taking the Poisson errors in each bin ($\sqrt{N}$) into account, and identify the approximate mass of the IMF break between them. Fig.~\ref{fig:IndividualIMF} shows examples of some of these power-law fits. The systematic uncertainly range in the power-law index of each Galactic component is estimated conservatively from the union of indices among all of the alternative samples for a given component.

\begin{figure}
    \includegraphics[width=\columnwidth]{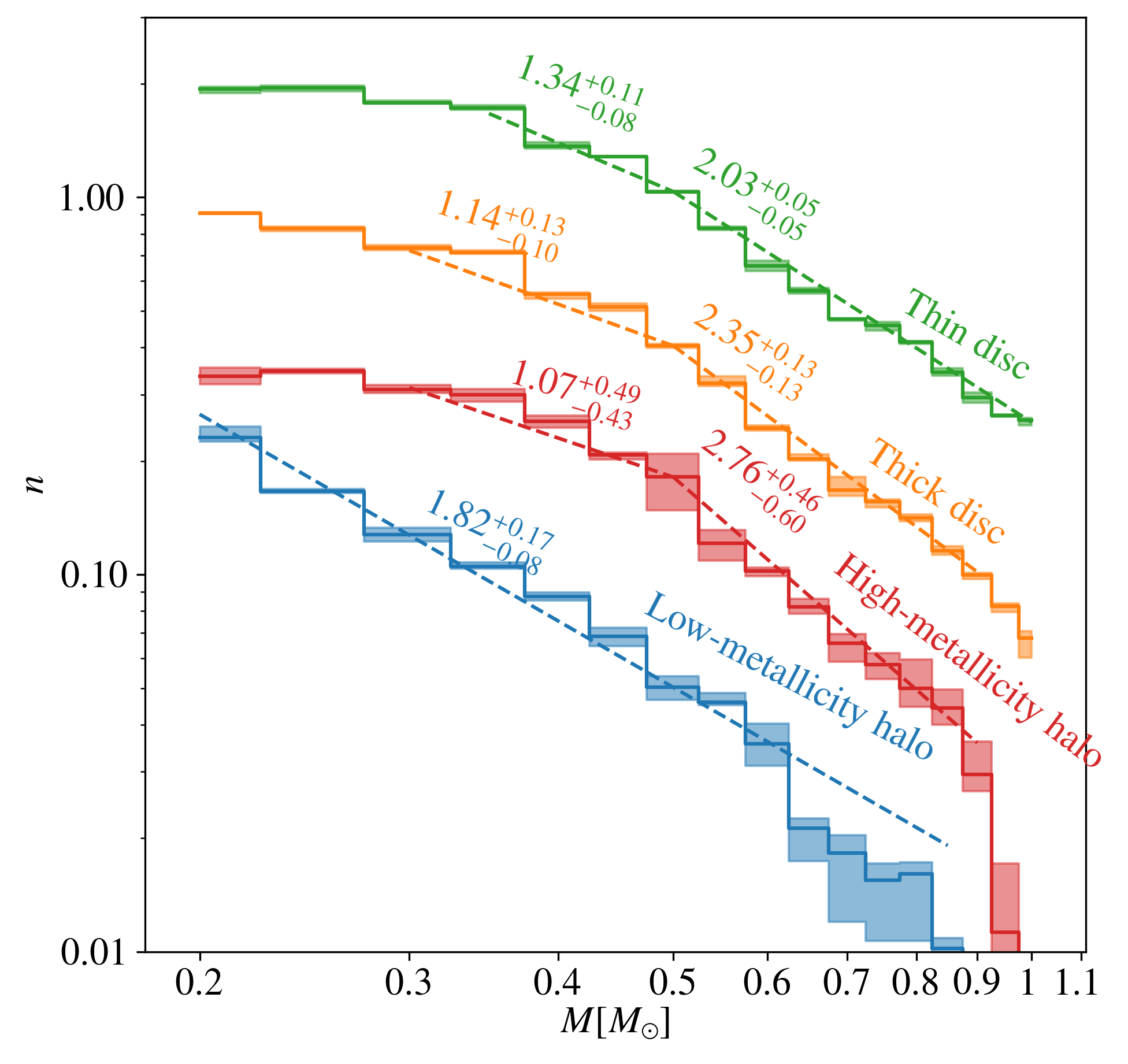}
    \caption{Normalized IMFs measured for each of the Galactic components, shifted from each other vertically for the sake of clarity. The central values are based on the full extinction- and binarity-corrected subsamples, within 100\,pc for the disc samples, and 250\,pc for the halo samples. The shaded regions mark the uncertainty of each IMF, based on the extreme values measured for all of the binarity-corrected subsamples of each component at the same distance. Tabulated data for this figure are given in Table~\ref{tab:MeanIMFs}.}
    \label{fig:IMF}
\end{figure}

Fig.~\ref{fig:IMF} compares the final IMFs for the four Galactic components, each with a band formed by the union of its alternative subsamples (within 100\,pc for the disc subsamples, and 250\,pc for the halo subsamples), representing the systematic uncertainty in the IMF. Tabulated data of this figure are presented in Table~\ref{tab:MeanIMFs}. Table~\ref{tab:IMF} summarizes the best-fit power-law indices and their uncertainties for each component, taking into account all binarity-corrected subsamples with more than 1000 stars (including subsamples with different volumes). Above $\sim 0.5$\msun, all of the IMFs are well described by steep power law of index $\sim 2$. Below 0.5\,\msun, the thin-disc, thick-disc, and high-metallicity-halo IMFs all turn over to a shallower slope. This region of the IMF can be fitted with a (shallower) power-law too, but given the further, incompleteness-driven flattening of the measured IMF at only slightly lower masses for these samples, it is hard to say whether or not a power-law is the appropriate functional form below the 0.5\,\msun\ break. The low-metallicity-halo IMF, in contrast, shows no evidence for any break or turnover, and is well-described by a single power law across the full mass range probed. It is ``bottom-heavy'' and distinct, in this respect, from the IMFs of the other components. This is the main result of this work.

\begin{figure}
    \includegraphics[width=\columnwidth]{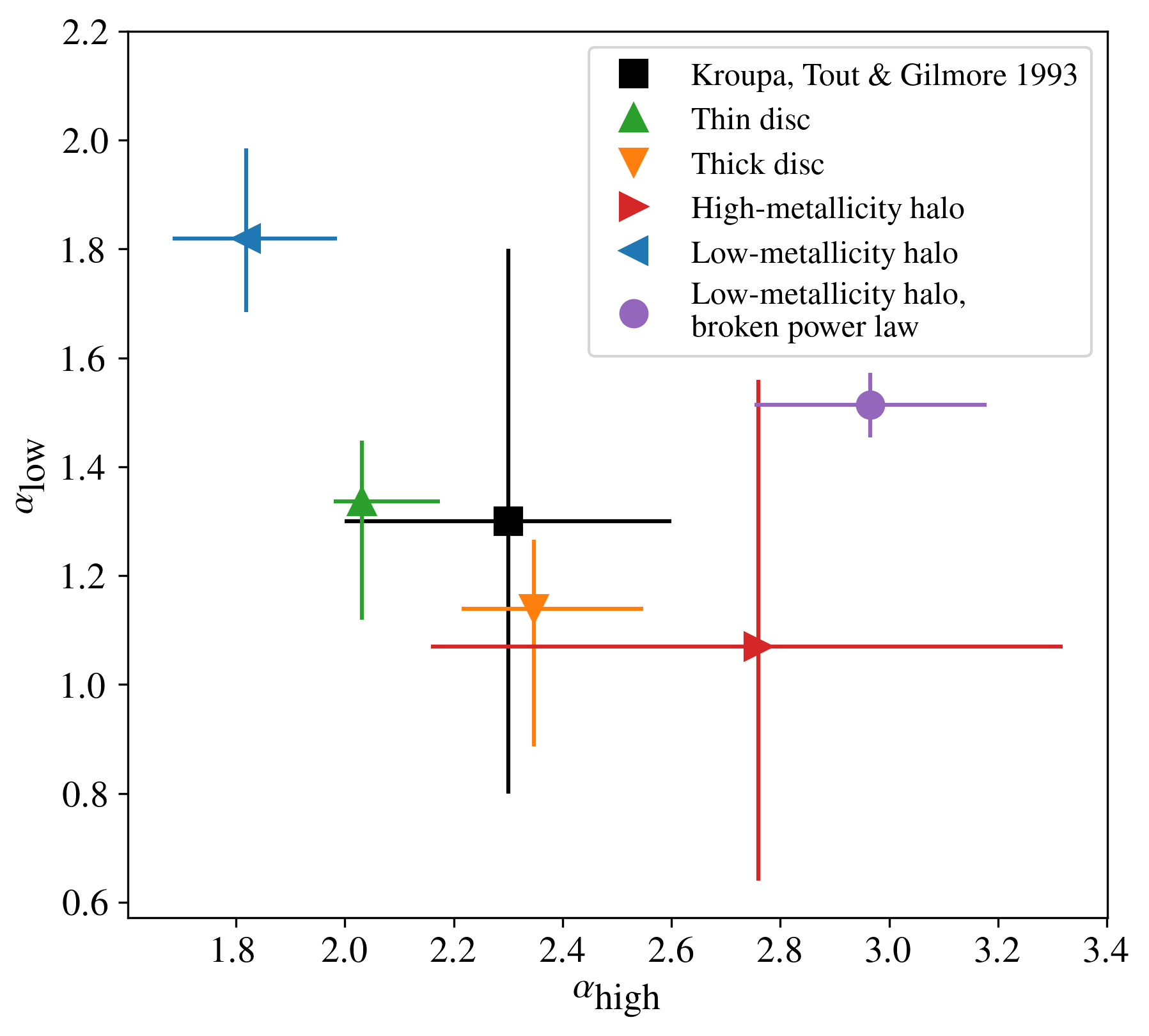}
    \caption{Comparison of IMF power-law indices below ($\alpha_\textrm{low}$) and above ($\alpha_\textrm{high}$) the break at $0.5$\,\msun, from Table~\ref{tab:IMF} and subsample ID \#48 in Table~\ref{tab:subsamples}, as indicated.
    The IMF indices of \citet{Kroupa_1993} as formulated in \citet{Kroupa_2001} are also shown for reference.
    }
    \label{fig:Indices}
\end{figure}

In terms of the actual slopes of the power laws representing the IMFs, the shallow low-mass slope of the IMF appears mutually consistent among the subsamples that display a 0.5-\msun\ break in the IMF, with slopes similar to the $\alpha_\textrm{low}=1.3$ power-law index of the popular \citet{Kroupa_2001} IMF, below its own 0.5-\msun\ break. Above the break, the IMF steepens slightly when going from the thin disc sample ($\alpha_\textrm{high}=2.03^{+0.14}_{-0.05}$) to the thick disc ($\alpha_\textrm{high}=2.35^{+0.20}_{-0.13}$) and to the high-metallicity halo ($\alpha_\textrm{high}=2.76^{+0.56}_{-0.60}$). The thin-disc high-end IMF slope is not far from, and yet significantly shallower than, that of the \citet{Kroupa_2001} value of $\alpha_\textrm{high}=2.3$. The low-metallicity halo, as noted, has no break, and is well described with a single power law of index $\alpha=1.82^{+0.17}_{-0.14}$. In other words, the single IMF slope we find for the low-metallicity halo is consistent with the high-mass-end slope of all the other components, and also of the \citet{Kroupa_2001} IMF. Fig.~\ref{fig:Indices} compares the power-law indices of the different Galactic components with the ``canonical'' IMF of \citet{Kroupa_1993}, as formulated in \citet{Kroupa_2001}.

\begin{figure}
    \includegraphics[width=\columnwidth]{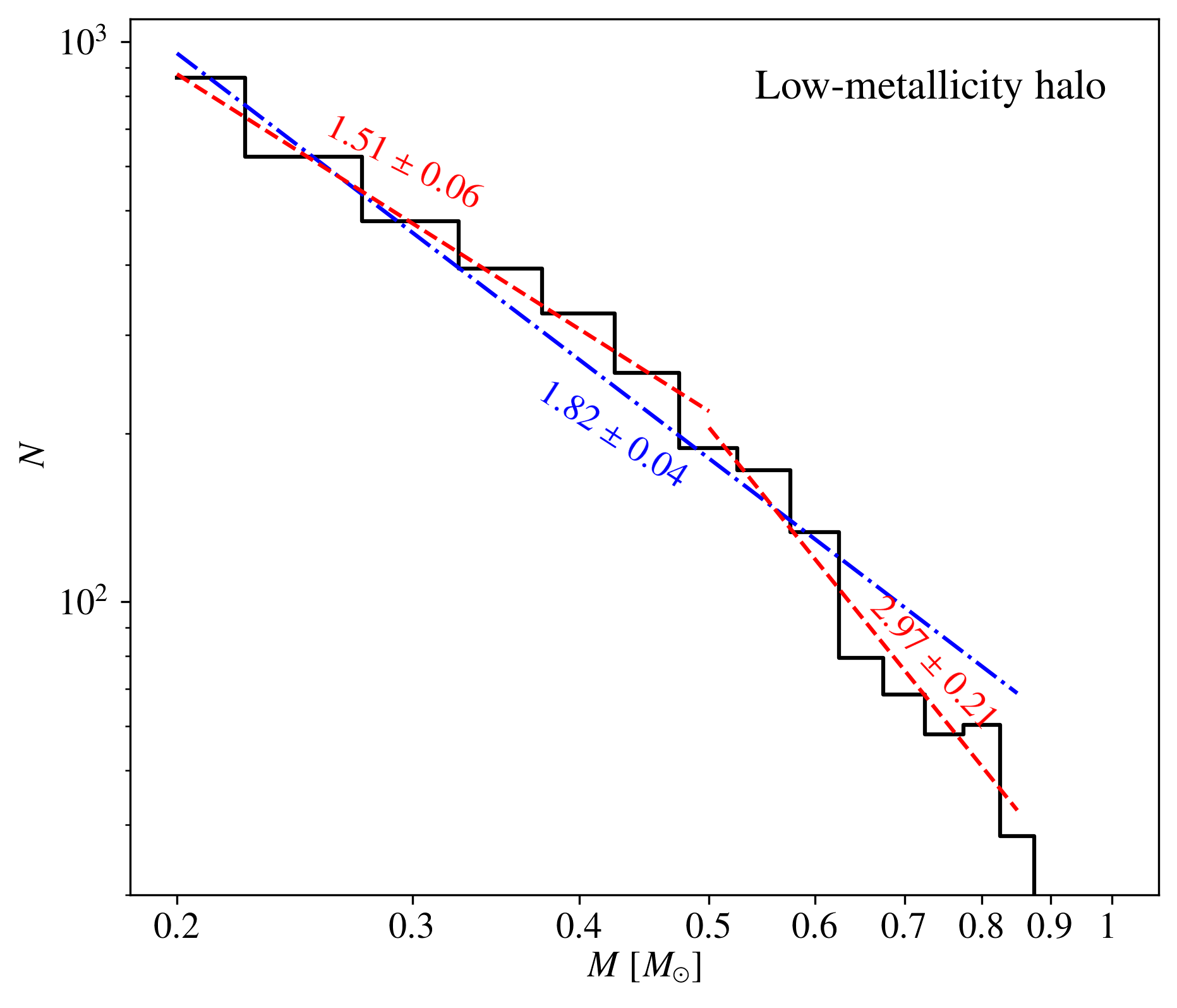}
    \caption{Measured IMF for the extinction- and binarity-corrected 250\,pc low-metallicity halo subsample (subsample IDs \#47-48 in Table~\ref{tab:subsamples}, solid black histogram), and comparison of a single power-law fit (dash-dotted blue) to  a broken power-law fit (dashed red).
    }
    \label{fig:LowZ}
\end{figure}

We have also explored how well a broken power law, with a break at 0.5\,\msun\,, can describe  the low-metallicity halo. A comparison between the single and broken power-law fits (subsample IDs \#47-48 in Table~\ref{tab:subsamples}) is shown in Fig.~\ref{fig:LowZ}. A broken power law indeed appears to fit the data ast least as well as the single power law but, either way, the resulting IMF is more bottom-heavy compared to that of the other Galactic components and to the canonical IMF (see Fig.~\ref{fig:Indices}).

To summarize, our work shows that Galactic stellar populations are distinguished by their IMF shape into three groups: thin-disc, with a broken power law IMF; thick-disc and high-metallicity-halo with a broken power-law of somewhat steeper high-mass-end slope; and low-metallicity-halo, with a single-power-law, bottom-heavy IMF, not unlike the \citet{Salpeter_1955} power-law IMF, with its $\alpha=2.35$.\footnote{ In fact, had Salpeter adopted the current age estimate for the Galactic disc (instead of the 6\,Gyr he assumed), he would have found $\alpha=2.05$ \citep{Kroupa_2019b}.}

\begin{table*}
    \caption{Normalized IMFs measured for each of the Galactic components, as in Fig.~\ref{fig:IMF}. The normalized number of stars in each mass bin is given in the $n$ columns; the lower and upper uncertainties are given in the $\sigma_{-}$ and $\sigma_{+}$ columns.}
    \label{tab:MeanIMFs}
    \centering
    \begin{tabular}{ccccccccccccc}
        \hline
         $M$ & \multicolumn{3}{c}{Thin disc} & \multicolumn{3}{c}{Thick disc} & \multicolumn{3}{c}{High-metallicity halo} & \multicolumn{3}{c}{Low-metallicity halo}\\
         (\msun) & $n$ & $\sigma_{-}$ & $\sigma_{+}$ & $n$ & $\sigma_{-}$ & $\sigma_{+}$ & $n$ & $\sigma_{-}$ & $\sigma_{+}$ & $n$ & $\sigma_{-}$ & $\sigma_{+}$\\
        \hline
        0.20 & 0.1240 & 0.0033 & 0.0018 & 0.1451 & 0.0002 & 0.0011 & 0.1339 & 0.0061 & 0.0073 & 0.2304 & 0.0055 & 0.0161 \\
        0.25 & 0.1247 & 0.0027 & 0.0017 & 0.1325 & 0.0024 & 0.0012 & 0.1384 & 0.0017 & 0.0021 & 0.1665 & 0.0018 & 0.0019 \\
        0.30 & 0.1140 & 0.0007 & 0.0014 & 0.1175 & 0.0019 & 0.0018 & 0.1239 & 0.0029 & 0.0035 & 0.1278 & 0.0054 & 0.0051 \\
        0.35 & 0.1102 & 0.0013 & 0.0021 & 0.1140 & 0.0007 & 0.0018 & 0.1196 & 0.0047 & 0.0045 & 0.1050 & 0.0013 & 0.0026 \\
        0.40 & 0.0874 & 0.0013 & 0.0021 & 0.0886 & 0.0026 & 0.0011 & 0.1021 & 0.0040 & 0.0034 & 0.0875 & 0.0021 & 0.0020 \\
        0.45 & 0.0820 & 0.0003 & 0.0004 & 0.0819 & 0.0018 & 0.0019 & 0.0831 & 0.0025 & 0.0012 & 0.0686 & 0.0041 & 0.0038 \\
        0.50 & 0.0662 & 0.0004 & 0.0005 & 0.0646 & 0.0007 & 0.0007 & 0.0725 & 0.0133 & 0.0111 & 0.0502 & 0.0038 & 0.0035 \\
        0.55 & 0.0529 & 0.0005 & 0.0005 & 0.0515 & 0.0009 & 0.0021 & 0.0484 & 0.0049 & 0.0041 & 0.0459 & 0.0009 & 0.0026 \\
        0.60 & 0.0420 & 0.0012 & 0.0014 & 0.0391 & 0.0006 & 0.0006 & 0.0409 & 0.0013 & 0.0008 & 0.0355 & 0.0045 & 0.0047 \\
        0.65 & 0.0361 & 0.0006 & 0.0006 & 0.0325 & 0.0003 & 0.0008 & 0.0329 & 0.0013 & 0.0016 & 0.0212 & 0.0038 & 0.0013 \\
        0.70 & 0.0304 & 0.0003 & 0.0002 & 0.0268 & 0.0009 & 0.0023 & 0.0263 & 0.0028 & 0.0016 & 0.0183 & 0.0062 & 0.0021 \\
        0.75 & 0.0293 & 0.0009 & 0.0006 & 0.0250 & 0.0009 & 0.0004 & 0.0231 & 0.0020 & 0.0017 & 0.0155 & 0.0048 & 0.0016 \\
        0.80 & 0.0265 & 0.0003 & 0.0001 & 0.0226 & 0.0005 & 0.0005 & 0.0199 & 0.0021 & 0.0039 & 0.0161 & 0.0054 & 0.0012 \\
        0.85 & 0.0220 & 0.0005 & 0.0005 & 0.0185 & 0.0004 & 0.0006 & 0.0177 & 0.0017 & 0.0021 & 0.0102 & 0.0022 & 0.0007 \\
        0.90 & 0.0188 & 0.0006 & 0.0006 & 0.0159 & 0.0004 & 0.0002 & 0.0118 & 0.0012 & 0.0026 & 0.0013 & 0.0006 & 0.0005 \\
        0.95 & 0.0169 & 0.0001 & 0.0001 & 0.0132 & 0.0005 & 0.0002 & 0.0045 & 0.0007 & 0.0024 & 0.0000 & 0.0000 & 0.0000 \\
        1.00 & 0.0164 & 0.0005 & 0.0003 & 0.0108 & 0.0012 & 0.0005 & 0.0009 & 0.0005 & 0.0004 & 0.0000 & 0.0000 & 0.0000 \\
        \hline
    \end{tabular}
\end{table*}

\begin{table}
    \caption{Combined IMF power-law indices. $\alpha_\textrm{low}$ ($\alpha_\textrm{high}$) is the IMF power-law index for masses lower (higher) than $\sim 0.5$\,\msun. The central values are based on the full extinction- and binarity-corrected subsamples (at a 100\,pc distance for the disc samples, and 250\,pc for the halo samples). The uncertainties are based on the extreme values (and the individual uncertainties) measured for all the binarity-corrected subsamples, having at least 1000 stars, of each Galactic component (at distances within 100 and 250\,pc for the disc samples, and 200 and 250\,pc for the halo samples).}
    \label{tab:IMF}
    \centering
    \renewcommand{\arraystretch}{1.5}
    \begin{tabular}{l c c}
        \hline
         Subsample & $\alpha_\textrm{low}$ & $\alpha_\textrm{high}$ \\
         \hline
        Thin disc & $1.34^{+0.11}_{-0.22}$ & $2.03^{+0.14}_{-0.05}$ \\
        Thick disc & $1.14^{+0.13}_{-0.25}$ & $2.35^{+0.20}_{-0.13}$ \\
        High-metallicity halo & $1.07^{+0.49}_{-0.43}$ & $2.76^{+0.56}_{-0.60}$ \\
        Low-metallicity halo & \multicolumn{2}{c}{$1.82^{+0.17}_{-0.14}$} \\
         \hline
    \end{tabular}
\end{table}

\subsection{Comparison to previous work}
\label{sec:PreviousWork}
The IMF that we have derived for thin-disc stars, which dominate in numbers the samples that we have analysed ($\sim 77$~per cent of the stars), is similar in form to the ``standard'' Galactic IMFs measured over the past decades \citep[e.g.][]{Zoccali_2000, Kroupa_2001, Chabrier_2003}, and until recently believed by many to be universal of star formation everywhere. For example, as already noted, in the sub-solar mass range that we have probed, the popular \citet{Kroupa_2001} IMF is described by a broken power law with slope $2.3$ above $0.5$\,\msun\ and $1.3$ below it. The thin-disc IMF that we have found has a similar shape and break position, with slopes of $2.03^{+0.14}_{-0.05}$ and $1.34^{+0.11}_{-0.22}$ above and below the break, respectively (see Fig.~\ref{fig:Indices} for a comparison). Considering the large number of stars, the completeness, and the precision of the \textit{Gaia} sample, as well as the straightforward derivation of the IMF for this mass range, we believe our result is among the more accurate and precise ones to date. However, to our knowledge, there have not been previous measurements of the local stellar IMF, separated by Galactic components, to which we can compare our results.

An interesting comparison we can make, nonetheless, is to the work of \citet{Sollima_2019} who, like us, used \textit{Gaia} DR2 to measure the IMF within our 50\,pc neighbourhood for the sub-solar mass regime, but for all such stars, rather than separated by kinematic components, as we have done. \citet{Sollima_2019} used a forward-modelling approach that begins from an assumed IMF and then uses stellar evolution models (PARSEC, as we have used, but mainly MESA Isochrones and Stellar Tracks \citep[MIST;][]{Choi_2016}) to predict the number density of stars as a function of position on the HRD. The IMF model is optimised to best fit the observed HRD. In contrast, the approach we have taken is to simply assign mass probability distributions to all the individual stars in the HRD, based on comparison  of the HRD position and the photometric uncertainties of each star to the PARSEC models. The stars are then divided among mass bins and a binarity correction is applied, to obtain the IMF. 

In order to compare our results to \citet{Sollima_2019}, we have defined an additional subsample (labelled ``All stars''), within the same volume (distance $<50$\,pc) as \citet{Sollima_2019}, but without kinematic distinctions, and have analysed it like our other samples, to derive the IMF. As noted, our correction for extinction is the same as that of \citet{Sollima_2019}. His treatment of binarity is different: rather than the correction we applied to the observed IMF to account for binarity (Eq.~\ref{eq:binary}), \citet{Sollima_2019} included binarity as part of the forward modelling process, but in a more simplistic way than ours---he assumed various fixed (i.e. mass-independent) values for the binary fraction $f_\textrm{bin}$, and a flat binary-component mass-ratio distribution, $P(q)$. Irrespective, both \citet{Sollima_2019} and we find that accounting for binarity has only minor impact on the derived form of the IMF in the mass range considered here ($M>0.3$\,\msun). 

\begin{figure}
    \includegraphics[width=\columnwidth]{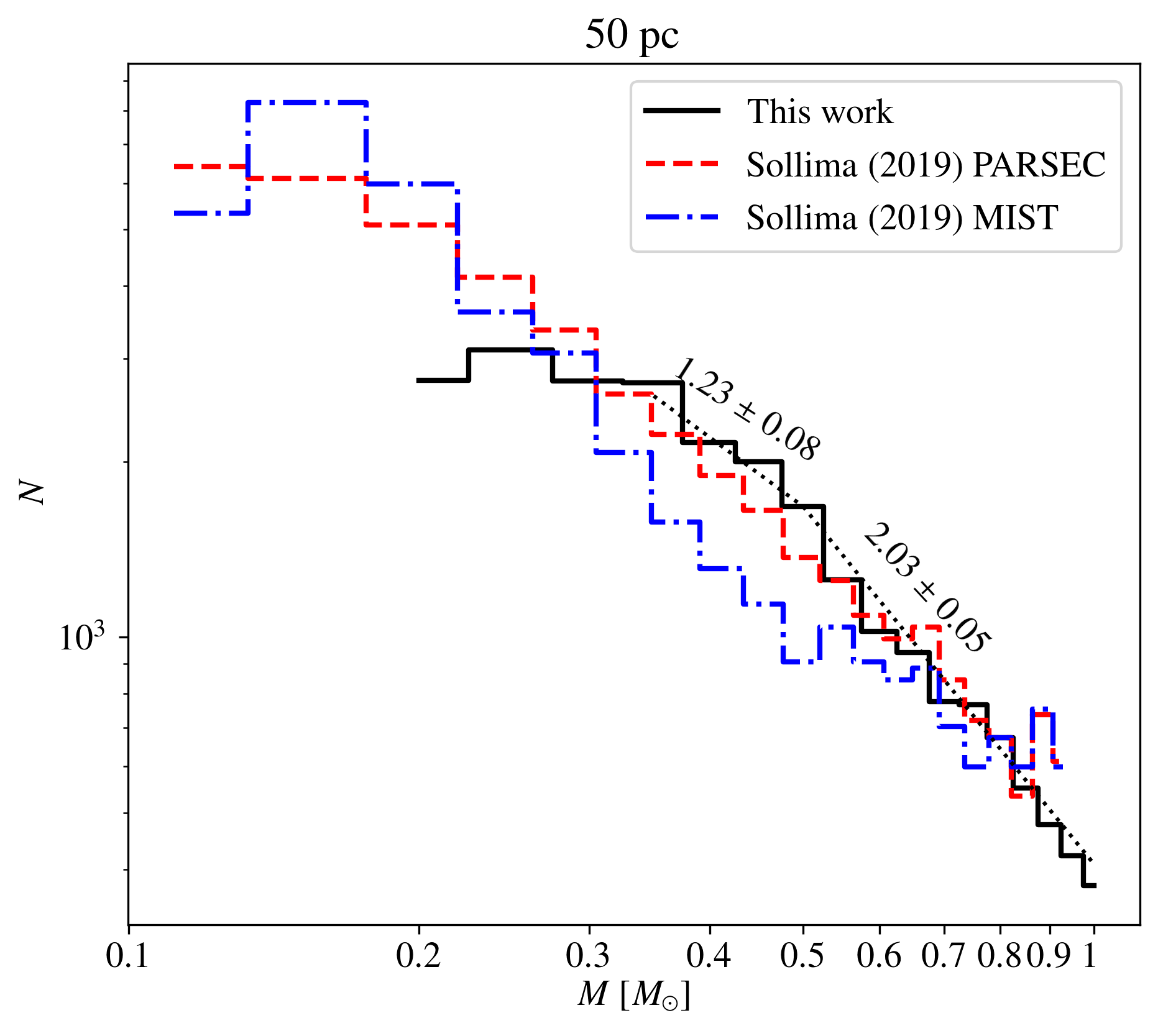}
    \caption{Comparison of our results to those of \citet{Sollima_2019}. The figure shows the IMF measured for our extinction-, twins-, and binarity-corrected 50\,pc ``All-stars'' subsample (solid black), and the fitted-IMFs of \citet[table~1]{Sollima_2019} using the PARSEC (dashed red) and MIST (dash-dotted blue) stellar evolution models, scaled to fit our measurements. The black dotted line shows the power-law fit to our IMF for this sample. See \S\ref{sec:PreviousWork} for details.}
    \label{fig:Sollima}
\end{figure}

Fig.~\ref{fig:Sollima} shows our ``All-stars'' sample IMF, compared to the IMF found by  \citeauthor{Sollima_2019}, as reproduced based on his table~1. The IMF we find for the 50\,pc ``All-stars'' subsample is, unsurprisingly, generally similar to that of our thin disc samples: a broken power law with index $\alpha_\textrm{low}=1.31 \pm 0.08$ in the range $0.35-0.5$\,\msun, and $\alpha_\textrm{high}=2.01 \pm 0.05$ in the range $0.5-1$\,\msun. Our IMF continues to flatten out in the $0.2-0.3$\,\msun\ range, presumably due to incompleteness to faint red stars, as already noted.

As seen in Fig.~\ref{fig:Sollima}, there appears to be good correspondence between our IMF and the IMF by \citeauthor{Sollima_2019} that is based (like ours) on PARSEC models, in the $0.3-0.8$\,\msun\ range. His models diverge from ours at higher and lower masses. The MIST-based model that was used by \citeauthor{Sollima_2019} for both the low and the high-mass ranges, is markedly different from ours, throughout the mass range considered here. \citeauthor{Sollima_2019}, considering this MIST-based result, argued for an IMF described by a single power-law of index $1.34 \pm 0.07$ in the range $0.25-1$\,\msun, and noted its departure from traditional IMFs that steepen above $\sim 0.5$ \msun, rather than only above 1\,\msun, as he finds. From Fig.~\ref{fig:Sollima}, however, it appears that his MIST-based IMF does not resemble a single power law. If anything, it looks like an inverted, or ``concave'', broken power law, with a steep low-mass slope and a shallow high-mass slope. \citet{Sollima_2019} measured, also with \textit{Gaia} DR2 and his same procedure, a similar IMF for the Pleiades star cluster. We believe the PARSEC models are better suited for IMF work, as the radii of stars in them have been empirically re-calibrated to avoid the radius inflation problem \citep{Morrell_2019}. The differences at the edges of the mass range between the IMFs we and \citet{Sollima_2019} have derived from the same data and with the same PARSEC models could be the result of the different completeness and binary treatment at the low-mass range ($M<0.3$\,\msun), and evolutionary effects at the high-mass range ($M>0.8$\,\msun). We argue, however, that our result is the more robust one, given the directness of our analysis method and the consistency of our results with many previous estimates of the IMF of the dominant disc population.

\subsection{Sensitivity of results to stellar-model uncertainties}

We have also explored the possibility that the steep low-mass IMF of the blue-halo stars (as opposed to the break to a shallower slope at low masses for the other Galactic components), rather than being a real effect, results from inaccuracies in the stellar models for this population. The blue halo stars are rare and of very low metallicities, and as such their model-predicted temperatures and radii, for a given stellar mass, have not been tested or calibrated, e.g., by means of stars of this type that are in eclipsing binary systems permitting mass measurement. As one possible example, supposing stellar models for this population over-predicted the stellar radii  of the low-mass stars, then the faint stars would be assigned erroneously low masses. This could shift stars to lower-mass bins in the IMF, washing out an intrinsic  break in the IMF, steepening its low-mass end, and thus artificially giving the IMF its observed form.

To assess this possibility quantitatively, we have experimented in artificially shifting by a fixed magnitude difference the $M_G$ values of all the blue-halo stars fainter than $M_G>8.5$\,mag (corresponding to masses $\lesssim 0.45$\,\msun). We find that a shift of at least $\Delta M_G\approx 0.2$\,mag is required in order to produce a distinct break in the IMF at $\sim 0.5$\,\msun\ and a low-mass IMF slope comparable to that of the other components. Therefore, for our result to be a consequence of this type of model inadequacy, stellar models of very-low-metallicity and low-mass stars would need to systematically over-predict stellar radii by $\sim 10$~per cent. We are not in a position to comment on whether or not a model error in this direction and of this magnitude is plausible. However, we make the following notes. The implied model radius over-prediction or ``observed radius deflation'' would be opposite in its sense to the known (but unexplained) phenomenon of ``radius inflation'', the under-prediction of the radii of about-solar-metallicity low-mass stars \citep{Morrell_2019}. Radius inflation has a magnitude of $3-7$~per cent. The PARSEC models that we use are re-calibrated to essentially eliminate radius inflation, so if deflation was behind the shape of our blue-halo IMF, it would imply not only that such a re-calibration should not be applied to very-low-metallicity stars, but that a further deflation of $3-7$~per cent of the model radii is needed for such stars. \citet{Morrell_2019}, however, found no significant dependence of radius inflation on metallicity, over the  limited range that they studied ($-0.3<\textrm{[Fe/H]}<0.3$). In view of all this, we believe it is unlikely that model inaccuracies drive of our main result, but future observational tests of models of low-metallicity stars must confirm this. 

\section{Discussion and conclusions}

We have measured the IMF of local ($<250$\,pc) Milky Way stars in the sub-solar-mass regime, separated according to Galactic kinematic and metallicity components. We were motivated to do this by: (a) the availability of \textit{Gaia} DR2, which for the first time permits derivation of an accurate and precise IMF separated by Galactic component; (b) some similarities, e.g. high \af\ ratios, between Galactic halo stars and the stars in massive elliptical galaxies that, as recently shown, have a bottom-heavy IMF, distinct from the ``universal'' IMFs usually considered; this led us to speculate that halo stars perhaps also have a bottom-heavy IMF; (c) the recent indications that the ``blue'' low-metallicity halo identified by \textit{Gaia} possibly consists largely or entirely of the stars of an ancient galaxy or galaxies that were accreted by the Milky Way $\sim 10$\,Gyrs ago. This raised the possibility that we could diagnose a distinct IMF in this foreign body ingested by our Galaxy.

We have limited our IMF study to sub-solar masses, to avoid the model-dependent complications of accounting for stars that have evolved off the main sequence. We have further adopted the simple yet reliable approach of using stellar-evolution models to assign a mass probability distribution to every star in our samples, and then counting stars in mass bins. Finally, we have corrected our measured IMF for the (small) effects of unresolved binary systems, using as input the latest empirical knowledge about the binary population. To gauge the level of systematic uncertainties in our IMFs, we have repeated our analysis for many different stellar subsamples, each somewhat differently defined.

For three of the four Galactic populations that we have studied---the thin disc, the thick disc, and those halo stars that in the HRD are concentrated in the ``red'', or higher-metallicity (relative to the ``low-metallicity halo'') locus---we find IMFs that are similar to those measured in previous decades in many different environments. They can be described as broken power laws, with a break at $\sim 0.5$\,\msun, a power-law index of $\alpha_\textrm{low}\sim 1$ below the break, steepening to $\alpha_\textrm{high}\sim 2$ above the break. We have now measured these slopes more accurately than was previously possible, and found evidence for a small but significant difference in $\alpha_\textrm{high}$ between thin-disc stars, on the one hand, and thick-disc and metal-rich halo stars, on the other. The similarity in IMFs between the thick disc and the metal-rich halo is in line with the recent suggestion that these two components belong to one continuous population of common origin---an ancient Milky Way disc structure that was heated and stirred, $\sim 10$\,Gyrs ago, by a merger or mergers with external galaxies such as Gaia-Enceladus or similar \citep[e.g.][see \S\ref{sec:intro}]{Helmi_2018, DiMatteo_2019, Amarante_2020, Belokurov_2020}. This holds also  if the driving event was, instead, the putative encounter of the Milky Way with the Andromeda galaxy \citep{Zhao_2013, Banik_2018, Bilek_2018}.

True to our original suspicions, we measure for the ``blue'', low-metallicity, halo a distinct, bottom-heavy, IMF, representable by a single power law across the mass range we have probed, and very reminiscent of the IMFs that have been deduced in massive elliptical galaxies. Even if the IMF of the blue halo is represented by a broken (rather than a single) power law, it is still overall bottom heavy compared to the other components. Curiously, massive ellipticals, with their high \af\ ratios, are \textit{high} metallicity stellar systems (with mean metallicities clustered near the solar value; e.g. \citealt{Conroy_2014}). In contrast, we have measured an early-type-like bottom-heavy IMF in the \textit{low}-metallicity Milky Way stellar halo, defined by metallicities in the range $-2<\textrm{\mh}<-0.6$.
As already discussed in \S\ref{sec:intro}, the blue halo is supposed to be composed, at least partly, of the stellar debris of the Gaia-Enceladus-Sausage galaxy, or galaxies, that merged with the Milky Way $\sim 10$\,Gyrs ago \citep{Belokurov_2018, Helmi_2018}. Estimates vary for the pre-merger mass of the external galaxy, but most indicate a stellar mass in the range $10^{8.5-10}$\msun\ \citep[e.g.][]{Mackereth_2019, Feuillet_2020}, which is substantial yet smaller than that of massive ellipticals. The blue-halo stars are spread in a broad, perhaps bifurcated, pattern in the \af\ vs. [Fe/H] plane, very different from that of ellipticals, whose stars are clustered at solar [Fe/H] and high \af. Alternatively, the blue halo might have formed in-situ from a large population of embedded clusters \citep{Kroupa_2002, Baumgardt_2008}, again a different environment than that of massive ellipticals.

It is therefore unclear what, if any, is the physical variable common to these two diverse stellar environments---early-type galaxies and the Gaia-Enceladus pre-merger galaxy---and which led to similar bottom-heavy IMFs in both. The variable could be age and/or the duration of star-formation. At the time of the merger or encounter, 10\,Gyrs ago, that created the blue halo, early-type galaxies were almost fully formed, a result of a brief burst of star formation, followed by a passive fading of their stellar populations \citep{Andreon_2016, Maoz_2017, SalvadorRusinol_2019}. The higher final metallicities and \af\ of ellipticals, compared to the blue halo, are perhaps the result of somewhat longer-duration bursts, or of more efficient retention within the deep galaxy potentials of SN-enriched gas, and its recycling in subsequent cycles of star formation.

In any event, it is becoming undeniable that the IMF is far from universal, and that different galactic environments harbour different IMFs. Our results show that the blue halo---possibly the stellar debris of a foreign galaxy long-ago accreted by the Milky Way---is distinguishable as such by the genetic signature of those stars---a peculiar, bottom-heavy IMF. At the same time, two Galactic components already suspected to have a common Milky-Way origin---the thick disc and the red halo---are shown to indeed have the same IMFs. Finally, the thin disc of modern Milky Way stars has yet a third form of IMF, slightly yet significantly different from the other two IMF forms. These results raise the hope that future IMF measurements across environments may provide clues to understanding the physics and history of star and galaxy formation.

\section*{Acknowledgements}
We thank Luca Casagrande, Doron Kushnir, and the anonymous referees for valuable comments.
The research of NH is supported by a Benoziyo prize postdoctoral fellowship.
This work was supported by grants from the Israel Science Foundation (DM), the German Israeli Science Foundation (DM),
and the European Research Council (ERC) under the European Union's FP7 Programme, Grant No. 833031 (DM). 
This work has made use of data from the European Space Agency (ESA) mission \textit{Gaia}\footnote{\href{https://www.cosmos.esa.int/gaia}{https://www.cosmos.esa.int/gaia}}, processed by the \textit{Gaia} Data Processing and Analysis Consortium (DPAC)\footnote{\href{https://www.cosmos.esa.int/web/gaia/dpac/consortium}{https://www.cosmos.esa.int/web/gaia/dpac/consortium}}. Funding for the DPAC has been provided by national institutions, in particular the institutions participating in the \textit{Gaia} Multilateral Agreement.

This research made use of \textsc{astropy}\footnote{\href{http://www.astropy.org}{http://www.astropy.org}}, a community-developed core \textsc{python} package for Astronomy \citep{Astropy_2013, Astropy_2018}, \textsc{matplotlib} \citep{Hunter_2007}, \textsc{numpy} \citep{Numpy_2006, Numpy_2011}, \textsc{scipy} \citep{Virtanen_2020}, and \textsc{topcat} \citep{Taylor_2005}, a tool for operations on catalogues and tables.

\section*{Data Availability}
The data underlying this article were accessed from the \textit{Gaia} archive\footnote{\href{https://gea.esac.esa.int/archive/}{https://gea.esac.esa.int/archive/}}. An ADQL query for retrieving the data is provided in Appendix~\ref{sec:GaiaQuery}.

\section*{Code Availability}
The code used to derive the stellar parameters is provided in the \textsc{stam} (``Stellar-Track-based Assignment of Mass'') \textsc{python} package, available through \textsc{github}: \href{https://github.com/naamach/stam}{https://github.com/naamach/stam}.



\bibliographystyle{mnras}
\bibliography{GaiaIMF}



\appendix

\section{The initial \textit{Gaia} query}
\label{sec:GaiaQuery}

Our initial \textit{Gaia} sample was acquired using the following ADQL query:

\begin{verbatim}
SELECT *
FROM gaiadr2.gaia_source AS gaia
JOIN gaiadr2.ruwe AS ruwe
ON gaia.source_id = ruwe.source_id
WHERE gaia.parallax_over_error > 10
AND gaia.phot_g_mean_flux_over_error > 10
AND gaia.phot_bp_mean_flux_over_error > 10
AND gaia.phot_rp_mean_flux_over_error > 10
AND gaia.parallax >= 4
AND 1.0 + 0.015*power(gaia.bp_rp,2) 
    < gaia.phot_bp_rp_excess_factor
AND gaia.phot_bp_rp_excess_factor 
    < 1.3 + 0.06*power(gaia.bp_rp,2)
AND ruwe.ruwe <= 1.4
\end{verbatim}

\section{Sample properties}
\label{sec:SampleProperties}

\begin{table}
    \caption{Median metallicity measured for each 250\,pc extinction-corrected kinematic subsamples.}
    \label{tab:Metallicity}
    \centering
    \renewcommand{\arraystretch}{1.5}
    \begin{tabular}{l c}
        \hline
         Subsample & $\mhf$ \\
         \hline
            Thin disc & $0.12^{+0.30}_{-0.27}$ \\
            Thick disc & $0.07^{+0.31}_{-0.30}$ \\
            High-metallicity halo 100kms & $-0.08^{+0.30}_{-0.27}$ \\
            High-metallicity halo & $-0.31^{+0.25}_{-0.24}$ \\
            Low-metallicity halo 100kms & $-0.99^{+0.28}_{-0.31}$\\
            Low-metallicity halo & $-1.06^{+0.25}_{-0.28}$ \\
         \hline
    \end{tabular}
\end{table}

\begin{figure}
    \includegraphics[width=\columnwidth]{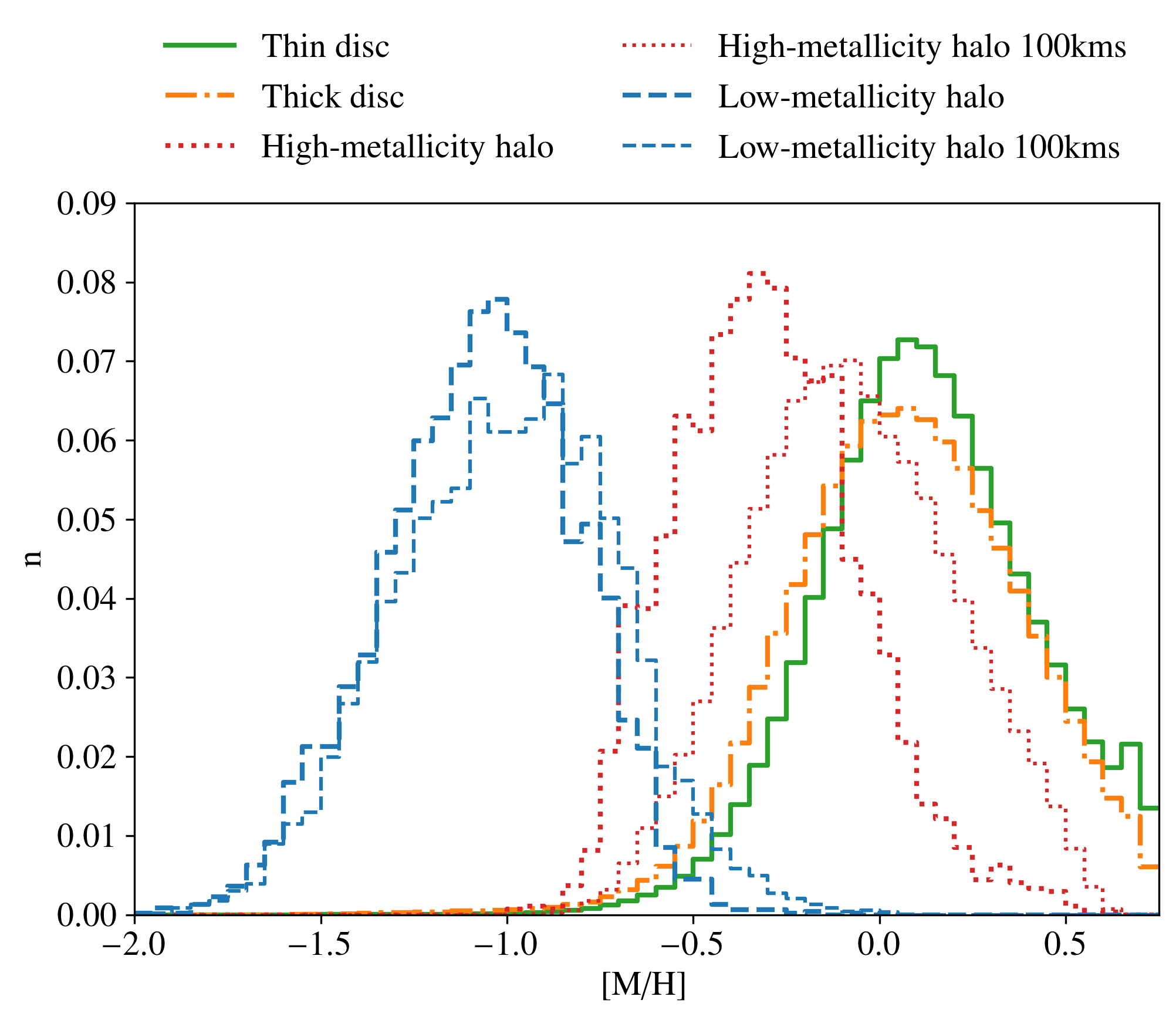}
    \caption{Normalized distribution of the mean metallicity measured for each star in the 250\,pc extinction-corrected samples, for each of the kinematic samples: thin disc (solid green), thick disc (dash-dotted orange), high-metallicity halo (dotted red), and low-metallicity halo (dashed blue). The $>100$\,\kms\ versions of the high- and low-metallicity halo samples are shown with thinner lines.}
    \label{fig:Metallicity}
\end{figure}

The metallicity distribution measured for each of the kinematic subsamples is shown in Fig.~\ref{fig:Metallicity}, and summarized in Table~\ref{tab:Metallicity}. The metallicity distributions of the thin and thick discs are seen to be similar. The red halo's metallicity distribution is shifted to lower values, but overlaps with that of the thick (and thin) discs. The red-halo subsamples's distribution, but with a lower transverse velocity threshold of 100\,\kms, is broader and has a peak at intermediate metallicity. This illustrates how the thick disc and the red halo can be viewed as one continuous population with a trend of lower metallicity with increasing velocity  (see also Fig.~\ref{fig:MHvsV}, below). The blue halo's low-peaked metallicity distribution, in contrast, does not depend on the velocity threshold. Note that the blue and red halos each have tails that extend beyond the $\textrm{\mh}=-0.6$ isochrone boundary that we used to define the two samples. This is a result of the \textit{Gaia} photometric and astrometric errors, which propagate into an uncertainty in the metallicity probability distribution of each star. 

The metallicity distributions for the blue and red halos are qualitatively similar to those in \citet{Sahlholdt_2019}, which were measured for red-giant stars within 2\,kpc (rather than main-sequence stars within 250\,pc, in the present work), with metallicities based on spectroscopically-calibrated multi-band photometry \citep[see also][]{Gallart_2019}. Our halo distributions, however, are shifted to systematically higher metallicities, by $\sim 0.2-0.4$\,dex. One explanation for this shift, or for part of it, could be our use of PARSEC isochrones with $\textrm{\af}=0$. The added atmospheric opacity resulting from, say, $\textrm{\af}=0.2$, which is reasonable for halo stars, is equivalent to an increase in \mh\ by $\sim 0.15$\,dex \citep{Salaris_1993}. An additional and comparable upward shift in the metallicity could be caused by an insufficient correction for extinction. Comparison of the metallicity distributions for our thin-disc subsamples, with and without extinction correction, shows a shift in \mh\ by about 0.03\,dex. If the recipe we have used for extinction underestimates the true extinction by a factor of a few, this could shift the distributions by $\sim 0.1$\,dex to higher \mh. Indeed, our thin-disc distribution peaks at $\textrm{\mh} \sim 0.1$, rather than nearer to zero, as one would expect. Finally, differences among the predictions of different stellar evolution models, based on their assumptions regarding element abundances and physical processes, may also lead to shifts of this order of magnitude in deduced metallicities \citep[e.g.][]{Serenelli_2017}.

\begin{figure}
    \includegraphics[width=\columnwidth]{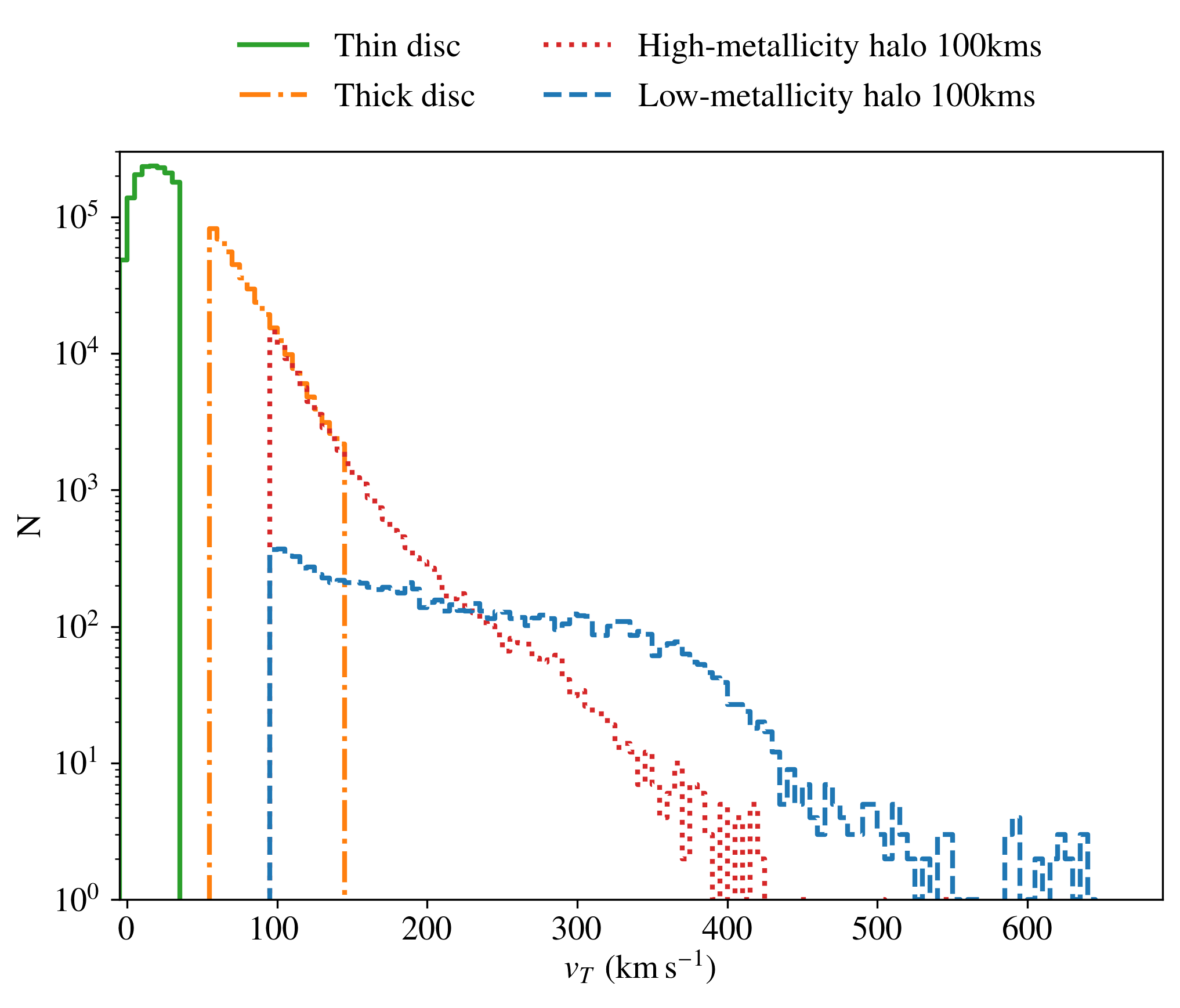}
    \caption{Transverse velocity distribution for each of the 250\,pc extinction-corrected kinematic samples: thin disc (solid green), thick disc (dash-dotted orange), high-metallicity halo (dotted red), and low-metallicity halo (dashed blue).}
    \label{fig:Velocity}
\end{figure}

Fig.~\ref{fig:Velocity} shows the transverse velocity distribution for each of the kinematic subsamples. The thin disc, the thick disc, and the high-metallicity halo, all appear to be parts of the same, roughly exponential, velocity distribution. The low-metallicity halo, in contrast, has an almost flat velocity distribution, up to $\sim 350$\,\kms\, where it cuts off exponentially.

\begin{figure}
    \includegraphics[width=\columnwidth]{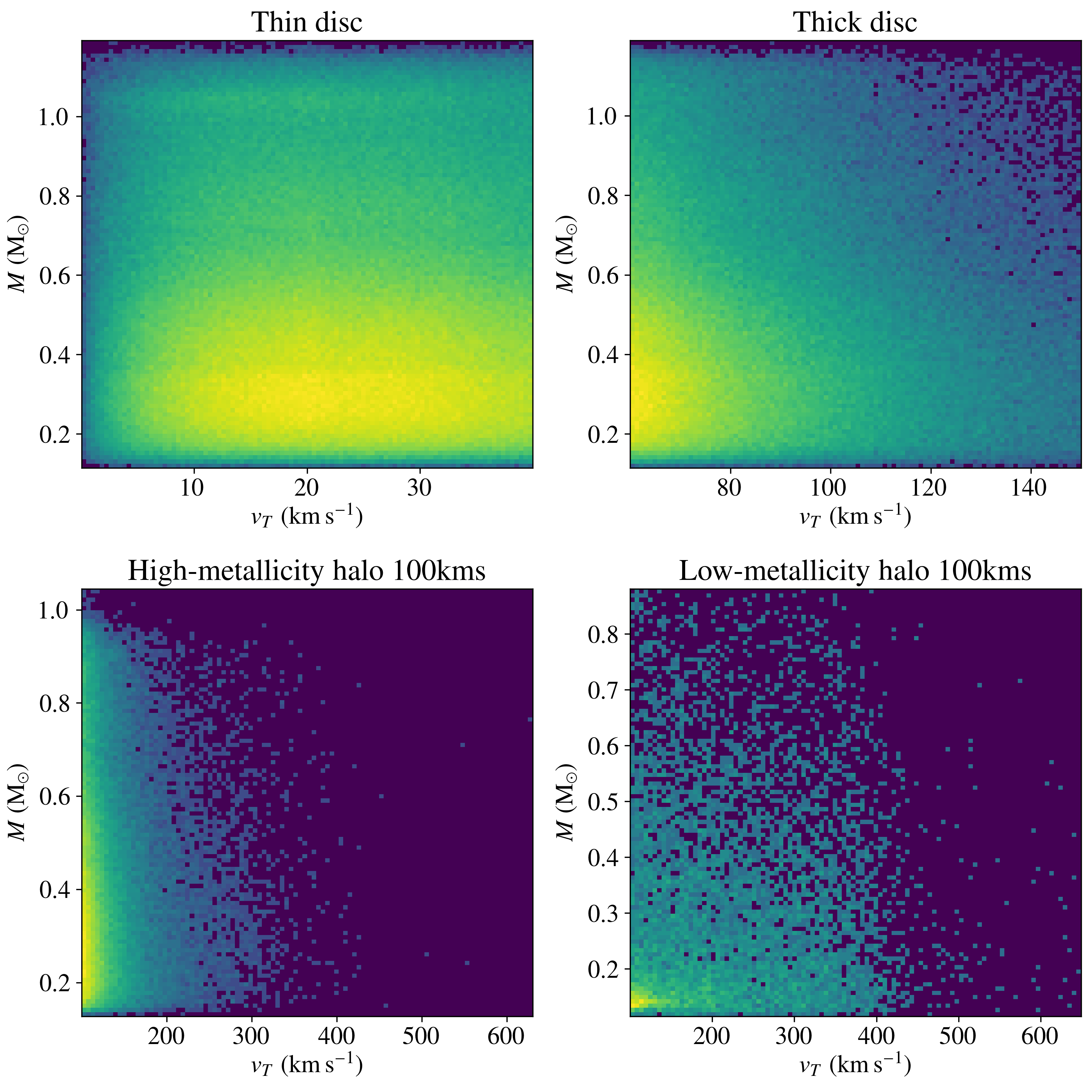}
    \caption{Two-dimensional histogram of the mean mass vs. transverse velocity for each of the kinematic subsamples.}
    \label{fig:MvsV}
\end{figure}

\begin{figure}
    \includegraphics[width=\columnwidth]{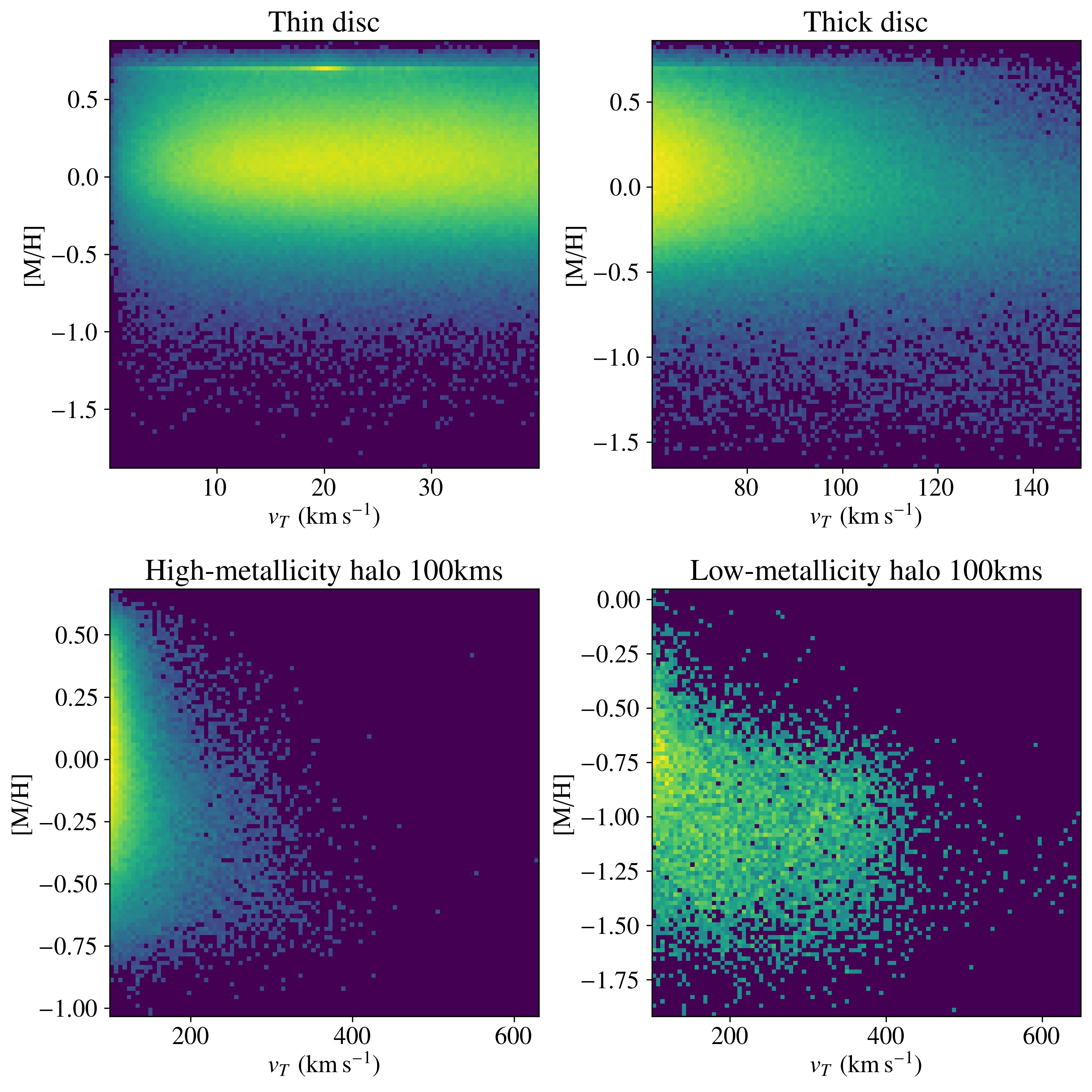}
    \caption{Metallicity vs. transverse velocity for each of the subsamples.}
    \label{fig:MHvsV}
\end{figure}

\begin{figure}
    \includegraphics[width=\columnwidth]{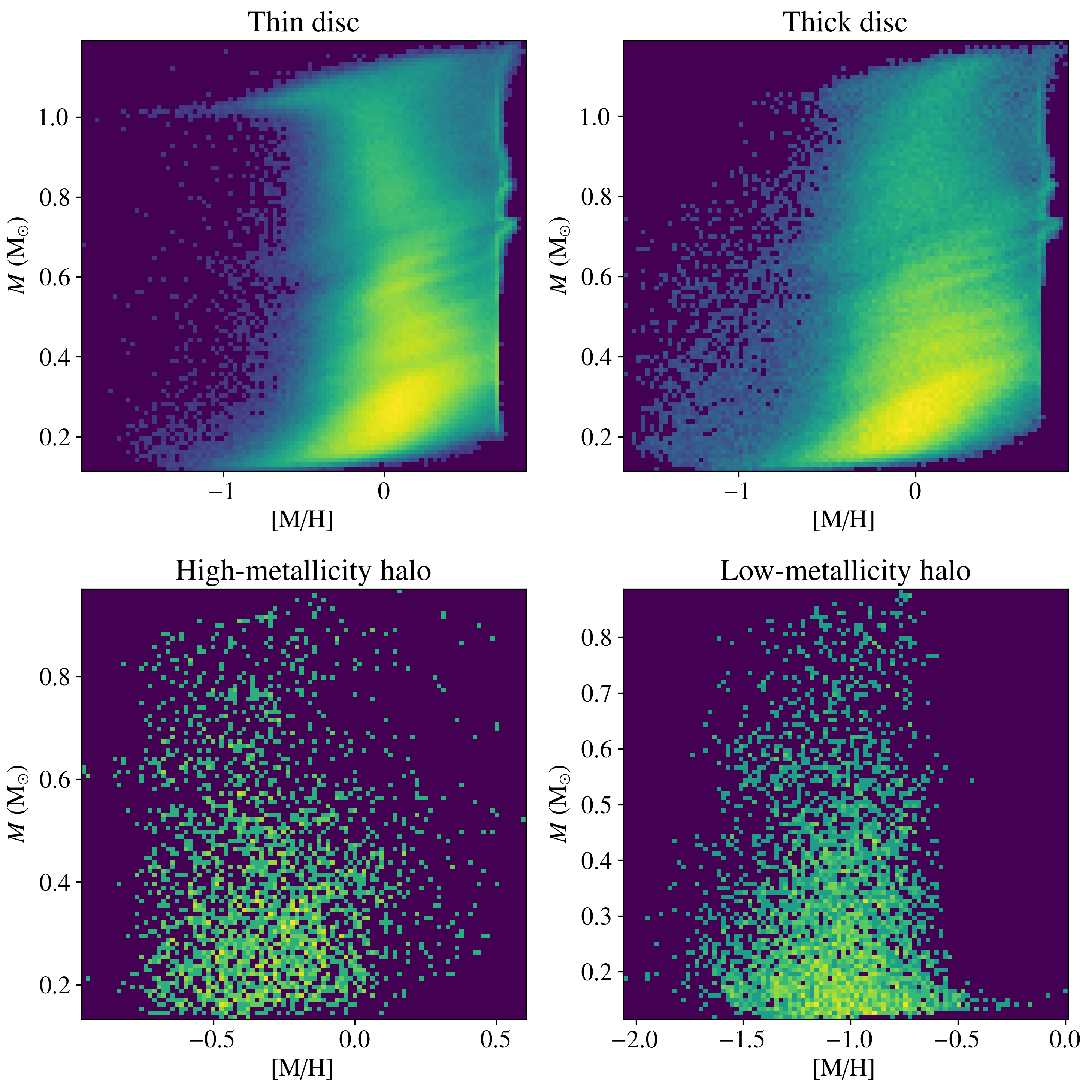}
    \caption{Stellar mass vs. metallicity for each of the subsamples.}
    \label{fig:MvsMH}
\end{figure}

The correlations between mass, metallicity, and velocity, are shown in Figs.~\ref{fig:MvsV}$-$\ref{fig:MvsMH}. Within each kinematic subsample, we see no significant correlation between mass and transverse velocity (Fig.~\ref{fig:MvsV}). For the thick disc and especially the red halo, however, one can see the trend of decreasing metallicity with increasing velocity (Fig.~\ref{fig:MHvsV}). There is no large-scale correlation between mass and metallicity, although some stripe-like substructures can be seen in Fig.~\ref{fig:MvsMH}. These sub-correlations resemble in form the isomass tracks used for the mass assignment, and result from the degeneracy in the location on the HRD between mass and metallicity.

\bsp	
\label{lastpage}
\end{document}